\documentclass[10pt,journal,compsoc]{IEEEtran}
\usepackage{times}
\usepackage{graphicx}
\usepackage{multirow}
\usepackage{hyperref}
\usepackage[section]{placeins}
\usepackage{longtable}
\usepackage{lipsum} 
\usepackage{blindtext}
\usepackage[dvipsnames]{xcolor}
\usepackage{rotating}
\usepackage{todonotes}
\usepackage{caption}
\usepackage{subcaption}
\usepackage{threeparttable}
\usepackage{amssymb}

\ifCLASSOPTIONcompsoc

  \usepackage[nocompress]{cite}
\else
  \usepackage{cite}
\fi

\ifCLASSINFOpdf

\else

\fi

\begin{document}

\title{ A Survey on Theorem Provers in Formal Methods}
\author{M. Saqib Nawaz, Moin Malik, Yi Li, Meng Sun and M. Ikram Ullah Lali
\IEEEcompsocitemizethanks{\IEEEcompsocthanksitem M. Saqib Nawaz is with School of Computer Science and Technology, Harbin Institute of Technology, Shenzhen, China.
\protect E-mail:msaqibnawaz@hit.edu.cn

\IEEEcompsocthanksitem Moin Malik is with  Department of Computer Science \& IT, University of Sargodha, Pakistan.
\protect  E-mail:moinmalik07@yahoo.com

\IEEEcompsocthanksitem Yi Li and Meng Sun are with School of Mathematical 
Sciences, Peking University, Beijing, China.
\protect E-mail:\{liyi\_math, sunm\}@pku.edu.cn

\IEEEcompsocthanksitem M. Ikram Ullah Lali is with Department of Computer Science, Faculty of Computing and IT, University of Gujrat, Pakistan.
\protect E-mail:ikramullah@uog.edu.pk}
}

\IEEEtitleabstractindextext{
\begin{abstract}
Mechanical reasoning is a key area of research that lies at the crossroads of mathematical logic and artificial intelligence. The main aim to 
develop mechanical reasoning systems (also known as theorem provers) was to enable mathematicians to prove theorems by computer 
programs. However, these tools evolved with time and now play vital role in the modeling and reasoning about complex and large-scale systems, 
especially safety-critical systems. Technically, mathematical formalisms and automated reasoning based-approaches are 
employed to perform inferences and to generate proofs in theorem provers. In literature, there is a shortage of comprehensive documents that 
can provide proper guidance about the preferences of theorem provers with respect to their designs, performances, logical frameworks, strengths,
differences and their application areas. In this work, more than 40 theorem provers are studied in detail and compared to 
present a comprehensive analysis and evaluation of these tools. Theorem provers are investigated based on various parameters, which includes: implementation 
architecture, logic and calculus used, library support, level of automation, programming paradigm, programming language, differences and application areas. 

\end{abstract}

}

\maketitle
\IEEEdisplaynontitleabstractindextext
\IEEEpeerreviewmaketitle
\IEEEraisesectionheading{\section{Introduction}\label{s1}}

Recent developments and evolution in Information and Communication Technology (ICT) have made computing systems more and more complex. 
The criteria that how much we can rely on these systems is based on their correctness. 
Bugs or any loopholes in the system lead to severe risks that endanger 
human safety or financial loss. In recent times, bugs ratio has 
increased due to the complex designs of the modern systems under market pressure and user requirements.
The efforts and cost required to correct bugs increases as the gap widens between their introduction and detection.
Table \ref{Ta1}, taken from \cite{White17}, shows the relative costs to fix bugs that are introduced in the requirements
phase. A bug introduced in a particular stage of a system development is relatively cheap to fix if also detected in that stage.
It becomes more hard and expensive to fix a bug that is introduced in one stage and detected in the other stage.
For safety-critical systems, the impact of bugs can be so large as to make a fix effectively mandatory.\vspace{-3mm}

\begin{table}[!ht]
 \caption{Cost to fix a bug introduced in requirements phase}\vspace{-2mm}
 \centering
\scalebox{0.8}{
  \begin{tabular}{ | c | c | }
    \hline
    {\bf Bug Found at Stage} & {\bf Relative Cost to Fix} \\ \hline
     Requirements & 1 (definition) \\ \hline
    Architecture & 3 \\ \hline
    Design & 5-10 \\\hline
    System Test & 10 \\\hline
   Post-Release & 10-100\\\hline
     \end{tabular}}
     \label{Ta1}
  \end{table}\vspace{-2mm}

Testing and verification techniques are used in the system test phase to empirically check their correctness. 
In testing, a system is tested against the software/hardware 
requirements \cite{Myers}. Similarly, simulation provides virtual environments for any real events. 
However, there are some inherent limitations of these techniques. A program can be only tested 
against the functional requirements of the system, which may be not refined and may contain ambiguities that leads to inadequate testing. 
Exhaustive testing of systems is not possible. Moreover, the time and budget constraints may affect the testing process. 
Simulations are also based on assumptions and does not always cover all the aspects of the system \cite{Davisd}. 
Furthermore, both testing and simulation can not be used efficiently for analyzing the continuous or hybrid systems. According to 
\cite{Dijkstra}: ``Program testing is an effective way to find errors, but it does not guarantee the absence of errors''.
On the other hand, formal methods formally verify the system correctness. 
Formal methods are ``mathematical-based techniques that are used in the modeling, analysis and verification of both the software and hardware 
systems'' \cite{Woodcock}. Formal methods allow the early introduction of models in the development life-cycle, against which the system
can be proved by using appropriate mathematics.
As mathematics is involved in the modeling and analysis, 100\% accuracy is guaranteed \cite{Hasan}.  But why we need 
formal methods in place of other well-known, widely acceptable and easy to use techniques such as testing and simulation? 
To answer this question, we first provide a few examples where testing and simulation failed.

Air France Flight 447 crashed in June 2009, which resulted into hundred of casualties. During investigation, it was found that the probe 
sensors were unable to measure the accurate speed of the plane, which provides the automatic disengagement of autopilot. Similarly, 
in August 2005, Malaysian Airbus 124 landed unexpectedly 18 minutes after taking off due to a fault in its air data inertial 
reference unit. There are two accelerators that control the airspeed of the flight, but one of them failed,  which resulted into a sudden 
rapid climbing and passed almost 4000 feet higher than expected without any warning. After investigation, it came to know that on 
the failure of the first accelerator, the second one used the falsy data of former due to input anomaly in the software. The probe, which laid hidden 
for a decade, was not found in testing because the designer had never assumed that such an event might occur \cite{Charette}. In June 
2009, Metro Train in Washington crashed and as a result, the operator of the train and 80 other people got injured severely. The cause of this 
incident 
was the design anomaly in the safety signal system. The safety system sent a green signal to the upcoming station, while the track was not empty. 
Similarly, other examples are failure of the London Ambulance Service's computer added dispatch system \cite{Finkelstein}, Therac 25 \cite{Leveson}, Anaesthetic 
equipment and the respiration monitoring device \cite{Mackie} which resulted into the  casualties and financial losses. All listed accidents could have
been avoided if the design of the systems were analyzed mathematically.

Formal methods techniques in contrast to testing permit the exhaustive investigation and reveals those bugs which are missed by testing methods. 
Actual requirements of the system in such techniques are 
translated into formal specifications which are mathematically verified and elaborate the actual behavior of the system in real scenarios. 
Two most popular formal verification methods are \texttt{model checking}
and  \texttt{theorem proving}. 
In  model checking, a finite model of the system is developed first, whose state space 
is then explored by the model checker to examine whether a desired property is satisfied in the model or not \cite{Baier}. Model checking is 
automatic, fast, effective and it can be used to check the partial specification of the system. However, model checkers 
still face the {\em state-space explosion problem} \cite{Clarke}. Model state space grows infinitely large with increase in the 
total number of variables and components used and the number of distinct values assigned to these variables \cite{Nawaz2}. 

Theorem proving on the other hand, can be used to handle infinite systems. In theorem proving, systems are defined and specified by users in an
appropriate mathematical logic. Important/critical properties of the system are verified by {\em theorem provers}. 
Theorem prover checks that whether a statement (goal) can be derived from a logical set of statements (axiom/hypothesis). It can 
model and verify any system that can be defined with the help of mathematical logics.
It is akin to Computer Algebra System (CAS) because both are used for symbolic computation. However, theorem provers have some advantages 
over CAS such as: flexibility in logic expressiveness, clear expression and more rigor. Theorem provers can be further categorized into 
two main types: Automated Theorem Provers (ATPs) and Interactive Theorem Provers (ITPs). ATPs deal with the development of automated computer 
programs to prove the goals \cite{Sutcliffe2}. In contrast, ITPs involve human 
interaction with computer in the process of proof searching and development. That is why ITPs are also known as {\em proof-assistants}. 
Due to practical limitations in pure automation, interactive proving is the suitable way 
for the formalization of ``most non trivial theorems in mathematics or computer system correctness'' \cite{Harrison2}.
Theorem provers have been used successfully in various domains such as biomedical \cite{Rashid}, 
game theory \cite{Kerber}, machine learning \cite{Kaliszyk2}, economy \cite{KLR16}, computer science \cite{Gallier}, 
artificial intelligence \cite{Wang} and self-adaptive systems \cite{Weyns12}.  
Note that the terms theorem provers and
mechanical reasoning systems are used interchangeably in this paper.

We believe that a comprehensive review on mechanical reasoning systems is strongly needed. People having little knowledge about them
generally think that all the systems based on mathematics have similar nature. However, it is not the case. Each system has different 
functionality and it is not an easy task to select which system should be used for the formalization efforts. 
Theorem provers are diverse in nature and the
main aim of this work is to demonstrate how different they are. Moreover, the goal is to provide a proper guidance to new 
researchers in formal verification. In order to substantiate our work, a questionnaire has been designed 
for the evaluation of theorem provers. The questionnaire is then filled by 
the developers and active researchers of theorem provers. Mechanical reasoning systems are investigated for the following parameters:

\begin{itemize}
\item Mathematical logic used in the system,
\item Implementation language of the system,
\item System type,
\item Platform-support in the system,
\item System category, whether it belongs to ATP or ITP,
\item Truth value of the system (binary/fuzzy),
\item Calculus (deductive/inductive) of the system,
\item Set (ZF/fuzzy) theoretic support in the system,
\item Programming paradigm of the system,
\item User Interface (UI) of the system,
\item Scalability of the system,
\item Distributed/multi-threaded,
\item Integrated development environment (IDE) support,
\item Library support in the the system,
\item Whether the system satisfy the de Bruijn criterion, and 
\item Whether the system satisfy the Poincar{\'e} principle of automation.
\end{itemize}

Primarily, this survey is a collection of tables and figures that illustrates various aspects of theorem provers. We 
received replies from experts/developers of 16 theorem provers. Another 27 theorem provers characteristics are investigated through 
online databases and research articles. 
We also report the top scientists who have proved maximum number of theorems in provers 
and top provers in which most number of mathematical theorems are proved till now. ATP that performed best at CADE ATP 
system competition (CASC) are also discussed. CASC is a yearly competition for the first-order logic fully ATPs. Moreover, aforementioned 
parameters are used to compare the provers and to present their main characteristics and differences. Finally, their applications 
are discussed along-with the existing recent work and the potential application/problems, where they can be used. 

The rest of the survey is structured as follows: An overview on the historical background of theorem provers in
the light of logical frameworks  is provided in Section \ref{s2}. Related work is also discussed. Section \ref{s3} elaborates the research 
methodology that is based on 
the systematic literature review in software engineering. Section \ref{s4} presents the results of the questionnaire, where answers of 
the experts and developers are presented. In Section \ref{s5}, top scientist that proved most of the theorems and top provers in which 
maximum number of theorems are proved till now are listed. ATP that performed best at CASC competition are also listed. Finally, theorem 
provers are compared for aforementioned criteria. The strengths, in-depth analysis of the differences among 
the existing theorem provers and their application areas are discussed in Section \ref{SAFD}, along-with the future research directions. 
The survey concludes with some remarks in Section \ref{s6}.

\section{Background} \label{s2}
In this section, historical background of the logical frameworks that are used in mechanical reasoning systems is presented. Furthermore, 
related work on the surveys and previous comparisons of theorem provers is discussed.

\subsection{Mathematical Logic}
Treating mathematics as a formal theory where all the mathematical statements are proved with a set of few basic axioms 
and inference rules is a long-standing goal. However, formal proofs of theorems require a lot of steps, effort and time. 
Many ancient Greek logicians, mathematicians and philosophers successfully 
expressed reasoning through syllogism \cite{Anellis}. Syllogism deals with formalization of deductive reasoning on logical arguments to 
arrive at a conclusion based on two or more propositions. Leibniz worked on the ways to reduce human reasoning calculations in symbolic 
logic and embodied these deductive reasoning into mathematics, making it easier to be 
implemented in computer programs. 
Mathematical reasoning provide objectivity and creativity that is hard to be found in any 
other fields. In mechanical reasoning, two paths were presented. One was to analyze human proof creation process and implement it using
computational resources. The other was to utilize the work of logicians and transform the logical reasoning into a standard form on which 
algorithms are based \cite{Gallier}. In the mid of 1950s, the relationship of the computer to mathematics has emerged in the form of 
automated reasoning, especially the automation of mathematical proofs \cite{Mackenzie}. More discussion on formal deduction,
LCF (Logic of Computable Functions) and modern type theory role in the development of theorem provers can be found at \cite{Maric, Barendregt01}.

Emergence of automated reasoning initiated the earliest work on computer-assisted proofs, when the first general purpose computers became 
available \cite{Harrison2}. The field of computer supported theorem proving gets attention in the second half of 20th century.
In 1970s, theorem provers were investigated for verification of computers systems. 
Extensive research in 
this area was done in late 1980s when these tools were used in the verification of hardware systems. In mid 1990s, a bug  
in Intel's Pentium II processor caused by floating point division increased the interest in formal methods and formal hardware verification tools 
were used by industry in their system design in late 1990s \cite{Kropf}. In 1994, Boyer proposed the QED manifesto \cite{Boyer} for a 
``computer-based database of all mathematical knowledge'' (formalized mathematics). During the years, the QED manifesto is adopted by many 
theorem provers. In recent years, the mechanical mathematical proofs for system verification has gained popularity \cite{Urban}. 
Future of formal methods looks promising 
and practical. Big companies such as Google, Facebook, IBM, Intel and Microsoft are now using and conducting research in formal methods. 

Theorem provers are fundamentally based on mathematical logics. Among the most popular theorem provers, we take three widely used ones: 
Propositional Logic, First-Order Logic (FOL) and Higher-Order Logic (HOL). Each logic is further discussed next.

\subsubsection{Propositional Logic}
Propositional logic is used to represent atomic propositions or declarative sentences with the help of mathematical Boolean operators such 
as \texttt{and, or, not, implication} and \texttt{equivalence}. It is also known as axiomatization of Boolean logic. Logical operators such as
 \texttt{conjunction} ($\wedge$), \texttt{disjunction} ($\vee$), \texttt{not} ($\neg$), \texttt{implication} ($\Longrightarrow$) and
\texttt{bi-implication} ($\Longleftrightarrow$) are used to bind propositions to make sentences. Truth values (\texttt{True} or 
\texttt{False}) are assigned to these propositions for evaluation of a sentence.  Axiomatization of Boolean algebra is also performed through
propositional logic. Strong argument about propositional logic is that it is decidable with the help of truth tables.

Newell developed the first theorem proving program {\em Logic Theorist} in 1956 \cite{Wrightson}. This program proved propositional logic theorems using
axioms/inference rules. It not only worked for numeric expressions but also for symbolic formulas and proof searching is guided through 
heuristics. It proved 38 theorems out of 52 and proofs of the theorems were more elegant. Another contribution in mechanical reasoning 
system was Davis and Putnam's semantics-based procedure reference \cite{Davis}. It was a decision method for checking whether a 
formula in Conjunctive Normal Form (CNF) is satisfiable or not. Such problems nowadays are called SAT (satisfiability) problems. They used ``ground
resolution'' for proving mathematical formulas in the form of predicate logic. Ground resolution used two propositional clauses and generate
another propositional clause. For testing Boolean formulas, Davis and Putnam's procedure implemented a series of ground resolution for 
proving satisfiability of these formulas \cite{Franco}. Satisfiability of Boolean formulas was also tested with Davis-Putnam-Longemann-Loveland 
(DPLL) method \cite{Logemann}. This was a searching algorithm based on backtracking mechanism. DPLL was used in checking the 
satisfiability of propositional logic formulas that are in CNF. This is an improved version of Davis and Putnam's procedure. This method used 
backtracking search instead of ground resolution. DPLL as compared to Davis and Putnam's method was much faster. It has a semantics searching 
process which helps in truth assignment. NP complete problem \cite{Fortnow} was the new notion, which was introduced several years later after 
DPLL. These problems were developed and proved by Cook \cite{Cook}. This motivated the development of SAT for solving hard problems, which has 
been improved over the last few years both on algorithmic and implementation level.

\subsubsection{First-Order Logic}
Propositional logic has less power as it is based on propositions and have no ability to predict any complex behavior.
Furthermore, it does not support variables, functions, relations and quantifications to compute complex problems. 
For example, ``Socrates is a man'' can be represented by propositional logic, but ``all men are mortal'' can not be 
represented in it because of quantification involvement. First-Order Logic (FOL) is the 
extension of propositional logic that allows quantifiers. Predicate logic is the general category to which FOL belongs. A predicate is a two 
valued or Boolean function that maps a given domain of objects onto a Boolean set, and this function is used to show specific quality or property 
between variables. For example, we assumed that \texttt{Q(n)} is a predicate where \texttt{n} is an even integer. Domain of discourse 
for this predicate is set of all integers. Therefore \texttt{Q(n)} depends on the value of \texttt{n}.  
Logical operators and quantifiers (\texttt{universal} ($\forall$), \texttt{existential} ($\exists$)) are used in predicate logic to express 
problems. It is less challenging to build automated theorem prover based on FOL. However, satisfiability checking of FOL is 
semi-decidable. Skolem and Herbrand first designed a semi-decidable procedure in 1920 \cite{Janicic}. Their method
was based on unguided retrieving for proof searching process and enumeration of ground terms. This method was useless in practical terms such as
proving non-trivial theorems. However, it played an important role for implementing theorem provers based on FOL. 

Prawitz was the first one who developed a general mechanical reasoning system for FOL \cite{Bibel}.
FOL provers such as Otter \cite{Mccune} and Setheo \cite{Letz} are mostly based on resolution and tableaux methods and they are used in 
solving puzzles, algebraic problems, software retrieval and verification of protocols. Some theorem provers even support recursive functions,
dependent and (co)inductive types, e.g. ACL2 \cite{Kaufman}, Metamath \cite{Megill} and E \cite{Schulz}. 
Such systems are used in mathematics (number theory and set theory), compiler verification, hardware verification and commercial applications.
However, exponential time algorithms are required for 
automatic proving in practice. Therefore, proofs in FOL are generally achieved by changing the FOL formula into a tautology 
or Boolean satisfiability problem. In such way, BDD, DPLL based SAT solvers can be used to automatically check the 
formulas. Satisfiability Modulo Theories (SMT) deals with the satisfiability of formulas against some logical theory \cite{Barrett09}.
In recent years, SMT solvers further extended the capabilities of SAT solvers.

\subsubsection{Higher-Order Logic}
FOL is more expressive as compared to propositional logic but less than Higher-Order Logic (HOL). FOL only quantifies variables. Predicates, 
propositions and functions are not quantified by FOL. For example:
quantifier quantifying a proposition
\begin{equation}\label{example}
    \exists s(R(y)\longrightarrow s)
\end{equation}

Another example is quantifier quantifying a predicate.
\begin{equation}\label{predicate}
    \forall S(Q(s)\wedge \neg R(s))
\end{equation}

HOL extends FOL by supporting many types of quantification. HOL permits predicates to 
accept premises which are also predicates, and allows quantification over predicates and functions which is not the case for FOL. Based on HOL,
one can construct a proof environment which is logically sound, do meta reasoning, interactive as well as automated and practically implementable. 
HOL is mostly used for ITPs. Various methods used in HOL-based theorem provers are decision
procedures, inductions, tableaux, rewriting, interactions and many heuristics. They are used in formalizing mathematics 
and verification of  programming languages, distributed system, compiler, software and hardware systems. 
Some well known HOL based theorem provers are: HOL \cite{Norrish}, Coq \cite{Bertot}, PVS \cite{R06a} and Agda \cite{Norell}. 

HOL is undecidable, so proving HOL properties is not fully-automatic and thus human assistance is required. It is more expressive and has the ability to prove 
complex problems and theorems. But it is more challenging to build an ATP  or ITP based on HOL. Generally, the proof process in ITPs is as follows. 
User first states the property or feature (in the form of a theorem) that is called a goal. User then applies proof
commands to solve it. Proof commands may decompose the goal into sub-goals. The proof process is completed once all the sub-goals are 
solved \cite{Yi}.

\subsection{Related Work}
Our survey on mechanical reasoning systems is certainly not the first one. 
Some work is done in the past on the survey and detailed discussion and comparison on theorem  provers 
\cite{Avigad, Harrison208, Maric, Hales, Woodcock, Harrison, Harrison2}. 
A detailed description of proof-assistants and their theoretical foundations is given in \cite{Barendregt01} along-with the comparison for nine theorem provers.
However, their comparison results take only one and a half pages. 
17 theorem provers are compared in \cite{Wiedijk} for three parameters: (i) library size of each prover, (ii) strength of 
the logic used in the prover, and (iii) level of automation in each prover. 
Similarly, \cite{Boldo} surveyed Coq, PVS, Mizar, ProofPower-Hol, HOL4, Isabelle/HOL and HOL Light for formally analyzing the  
real time systems. They also investigated the extended standard libraries 
that play main role in proof automations: C-CoRN/MathClasses for Coq, ACL2(r) for ACL2 
and the NASA PVS library. The application of theorem provers in economics is discussed in \cite{KLR16}, with the focus on two domains: 
social choice theory and auction theory. 

In literature, other comparisons between provers can be found. However, in most of those works, only two systems are compared generally. 
Such works include the comparisons between HOL and PVS \cite{Griffioen},
HOL and Isabelle \cite{Agerholm2}, NuPRL and Nqthm \cite{Basin}, Coq and HOL \cite{Jakubiec}, HOL and ALF \cite{Agerholm} and Isabelle/HOL and Coq \cite{Y18}. Some works have also been done on how to adapt the proofs to different systems \cite{Naumov, Obua06, Gordon4}. 
In \cite{Gastel}, Reentrant Readers Writers problem is first modeled in UPPAAL model checker and found a possible
deadlock scenario. They further converted the UPPAAL model and analyzed the model in PVS and checked the PVS model for arbitrary number of 
processes. Moreover, SPIN model checker is used in \cite{Gastel2} for modeling and analysis of 
Reentrant Reader Writers problem.
Promela model is converted to PVS specification and the correctness of the model was then verified. 

\section{Research Methodology}\label{s3}
Core part of a Systematic Literature Review (SLR) is the requirement of research questions. Right and meaningful questions are demanded 
to ask during the review process and we also need to pinpoint the scope of research accomplishments. Following the guidelines of 
\cite{Kitchenham2}, we have structured the research questions with the support of PIOC (Population, Intervention, Outcome, and Context) 
standard for applying the SLR process in software engineering field. PIOC for this work is presented in Table  \ref{modeltools}.
\vspace{-2mm}
\begin{table}[!htb]
 \caption{PIOC for this work}\vspace{-2mm} 
 \centering 
\begin{tabular}{|c|p{6.3cm}|}
  \hline
  \textbf{Population} & Mechanical reasoning systems\\  \hline
  \textbf{Intervention} & Theorem proving approach \\    \hline
   \textbf{Outcome} & Comprehensive document, evaluation and comparison \\       \hline
  \textbf{Context} & Developers and experts from industry and researchers from academia \\
  \hline
\end{tabular}
\label{modeltools}
\end{table}\vspace{-.2cm}

Efforts are made to collect evidence on recent scenarios of research in the development
of mechanical reasoning tools. For this purpose, we have designed research questions which are presented in Table
\ref{modeltools3}. 
We forwarded our questionnaire to a number of theorem prover developers, experts and forums.
Names and answers of the domain experts that responded are presented in Section \ref{s4}. \vspace{5mm}
\begin{table}[!ht]
 \caption{Research questions from our questionnaire}\vspace{-2mm} 
 \centering 
\begin{tabular}{|c|}
      \hline
{\bf Research questions about system general information}\\\hline
  What are the names of people who contributed into the system? \\ 
  When system first time (date, year) appeared in the market? \\ 
  What is the latest version of the system? \\ 
  When was the system updated last time? \\ 
  What is the address of web page for accessing it online? \\ 
  What are the unique features of the system? \\ 
  What are the success stories of the system?\\ 
      \hline
{\bf Research questions about system category information}\\\hline
  What is the type of the theorem prover (ATP/ITP)? \\ 
  What is the type of the system w.r.t. reasoning (mathematical)? \\ 
  What is the type of the system w.r.t. logic (FOL or HOL)? \\ 
  What is the type of the system w.r.t. truth values (binary/fuzzy)? \\
  What is the type of the system w.r.t. calculus (deductive/inductive)? \\
  Either set theoretic support (ZF/fuzzy) is available in the system? \\
  \hline

{\bf Research questions on system programming framework}\\\hline
  What is the paradigm (functional/imperative/other) of the system? \\ 
  What is the programming language (C/C++/Java/Other) of the system? \\ 
  What is the user interface (GUI/CLI) of the system? \\ 
  What is the scalability (distributed/multi-threaded) of the system? \\ 
  Either library support is available in the system? \\ \hline
\end{tabular}
\label{modeltools3}
\end{table}

\subsection{Keywords Retrieval for Search Strings}
Keywords (question elements) are find out from relevant research articles that we 
studied. These keywords are used for retrieving more information related to the development of mechanical reasoning tools from electronic databases.
These keywords are presented in Table \ref{modeltools55}. Alternative words for the keywords are 
find out by using synonyms and thesaurus. These alternative keywords are also used for the searching process. 
Keywords are linked together with Boolean \emph{OR} and search strings are constructed by linking the four \emph{OR} lists with Boolean \emph{AND}. 

\begin{table}[!ht]
 \caption{Keywords from mechanical reasoning surveys}\vspace*{-.2cm} 
 \centering 
\begin{tabular}{|p{1.2cm}|p{6.6cm}|}
  \hline
  \textbf{Armstrong et al.} \textbf{\cite{Armstrong}} & Formal tools, software and hardware correctness, provably correct design, 
  theorem provers, Satisfiability Modulo Theory, abstract models, Event-B, digital systems, formal methods, model correctness. \\   \hline
  \textbf{Mackenzie} \cite{Mackenzie} & Mathematical proofs, interactive theorem proving, automatic theorem proving,  mathematical logic, 
  mathematical reasoning, formal logic, proof automation, classical and constructive logic, machine intelligence. \\   \hline
  \textbf{Azurat \& Prasetya} \cite{Azurat} & Theorem prover, proof checker, formal verification, 
   programming logic, formal representation, HOL, PVS, automatic proof generation, Coq. \\   \hline
  \textbf{Harrison}\cite{Harrison} & Logic and program meaning, mathematical logic, symbolic manipulation, algebraic interpretation, formal languages, 
                            software engineering. \\     \hline
  \textbf{Harrison et al.}\cite{Harrison2} & Formal proof, interactive theorem provers, proof goals, semi automated mathematics, Automath, Coq, NuPRL, Agda, 
                              Logic of computable functions, HOL, PVS, proof language, proof automation.  \\     \hline
  \textbf{Boldo et al. }\cite{Boldo} & Proof assistant, formalization, proof libraries, interactive theorem provers, PVS, Coq, HOL4, Isabelle/HOL, 
                          ProofPower-HOL, HOL Light, proof automation. \\\hline
  \textbf{Wiedijk}\cite{Wiedijk} & Proof assistants, proof kernel, logical framework, decidable types, dependable types, de Bruijn criterion, 
                            Isabelle, Theorema, HOL, Coq, Metamath, PVS, Nuprl, Otter, Alfa, Mizar, ACL2. \\      \hline
  \textbf{Mari\'{c}}\cite{Maric} & Decision procedures, proof search, theorem provers, software correctness, interactive theorem provers survey, formal deduction, 
                    proof checking, logical frameworks, SAT solvers, SMT solvers, Poincar{\'e} principle. \\       \hline
\textbf{Hales}\cite{Hales} & Computer proofs, proof assistant, small proof kernel, logical framework, proof tactics, first-order automated 
                           reasoning, mathematical proof, theorem provers. \\       \hline
\textbf{Avigad \& Harrison}\cite{Avigad} & Axiomatic set theory, mathematical proof, calculus of reasoning, formalized mathematics, 
                                Formal verification, interactive theorem proving, Poincar{\'e} conjecture, formal proof systems. \\       \hline
\textbf{Barendregt \& Geuvers}\cite{Barendregt01} & proof checking, mathematical logic, type theory and type checking, type systems, predicate logic, higher-order logic
                                proof development, proof-assistants, Coq, Agda, NUPRL, HOL, Isabelle, Mizar, PVS, ACL2. \\       \hline                          
\end{tabular}
\label{modeltools55}
\end{table}

Online databases, journals and conferences related to mechanical reasoning tools are used for comparison, analysis and evaluation. 
Seven electronic sources of relevance in software engineering is identified in \cite{Brereton}.
However, in last few years, many new and famous libraries are developed especially in computer science field. Therefore,
it may also be necessary to consider other sources. The search strings were used on 10 digital libraries: 
 (i) DBLP, (ii) IEEE Explore, (iii) ACM Digital Library, (iv) Springer Link, (v) Science Direct, (vi) CiteSeer$^{X}$, (vii) Scopus, (viii) Inspec, 
 (ix) EI Compendex, and (x) Web of Science.

\section{Provers and Their Characteristics} \label{s4}
In this section, answers of the developers and experts that responded to our questionnaire are presented. The order in which we present the
answers of our respondents is the order in which we received their replies. In this way, we wish to express our gratitude to them.
\\\\
{\bf Matt Kaufman (ACL2)}: Matt Kaufman is a senior research scientist working at Department of Computer Science, University of Texas, Austin. Matt provided information about 
``ACL2'' \cite{Kaufman}. Main authors are M. Kaufman and J. Moore. However, 
several others have also made significant contributions. The first public release of ACL2 was 1.9 in 1994. The latest version is 8.2 and updated last time in 
May 2019. Basically, it is a monolithic system, but the applicative style of programming often makes it 
straightforward to use pieces of the system. It is generally classified as an ITP. However, it takes automation seriously; in that sense it 
shares characteristics with ATP. Its logic is FOL with induction and is written mostly in the ACL2 language, which is 
an applicative language extending a non-trivial subset of Common Lisp. Its UI is typically Emacs based. ACL2 is a cross
platform tool: it runs on Linux, MacOS and Windows. It also runs on the top of 6 different common Lisp implementations. Input format of ACL2 is 
{\em s-expressions}, though output can be pro-grammatically produced. Web address is \url{cs.utexas.edu/users/moore/acl2}.
ACL2 has been scaled to large applications, recently at Centaur and Oracle. Its users seem pretty happy with readability, but others might be 
put off by the \emph{s-expression} format. Inter-operability between different provers is limited, though there has been work 
\cite{Gordon4, Gordon5} that connects ACL2 and HOL4, e.g. ACL2 is first-order, but is still quite expressive because of its support for 
recursive definitions. Several capabilities allow it to do some things that might be considered higher-order in nature: macros, a proof technique 
called functional instantiation and oracle-apply. ACL2(p) \cite{Rager} supports parallelism for execution, proof and other infrastructure supports parallelism at the
level of collections of files. Run-time assertions are supported and lots of debugging tools are available for program execution and proof.
ACL2 can often emulate other logics by formalizing their proof theories. It is extensible or programmable by the user via rule classes and 
directly via meta rules and clause-processors. ACL2 may be the only ITP that presents a single logic for both its programming language  
(provide efficient execution) and its theorem prover (including definitions and theorems to prove). 
There is a large library of ``Community Books'' developed over many years by users, in daily use. There are users in academia,
government and industry.  
\\\\
{\bf Stephen Schulz (E)}: Stephen Schulz is the next person who responded to our questionnaire. Stephen designed and developed ``{\em E}'' theorem prover \cite{Schulz}. The first 
public release of \emph{E} was 0.2 in 1998. The latest version is 2.4 and updated last time in October 2019. \emph{E} was originally developed at 
TU Munich, but now it is maintained and extended at DHBW Stuttgart, Germany. License type is open source/free software under GNU GPL Version 2. 
It is an ATP for full FOL with equality, where first-order problems are reduced to clause normal form and uses a saturation procedure based on 
the equational superposition calculus. Main user community of the system is mathematician.
Web address is \url{www.eprover.org}.
\emph{E} won  several CADE ATP competition and has a good ranking. The type of \emph{E} with 
respect to reasoning is mathematical, type with respect to the logic is classical FOL and with respect to the truth value is binary.
Calculus used in \emph{E} is deductive and set theoretic support is available but on logical level via axiomatization for ZF. 
Programming paradigm is imperative, it is purely developed in C and it supports CLI (command line interface). \emph{E} is officially
distributed in source files and supports Linux, Mac OS, FreeBSD, Solaris, Windows and w/Cygwin. It is not a multi-threaded system, has a mixed
architecture (modular + monolithic) and has its own library. Proof can be generated in TPTP-3, PCL2 and Graphviz format. 
Some input codes are generated automatically from test data. System has no dedicated proof kernel, but has explicit proof-object. 
Any standard text editor can be used for files input. \emph{E} has been combined with other systems (Waldmeister, LEO-II, Vampire, Z3, etc.) at 
Isabelle Sledgehammer tool \cite{Paulson2} to increase the level of proof automation. 
\\\\
{\bf Makarius Wenzel (Isabelle)}: Makarius Wenzel provided information about ``Isabelle'' \cite{Paulson}, originally published by L. Paulson (Cambridge, UK). 
Many people have contributed to Isabelle in the last 30 years. It was released in 1986 and its pure logical framework first 
came up in 1989. Latest version of the system was released in June 2019. Isabelle grew out of 
university research projects, but it is of industrial quality, or even beyond that, because it is not subjected to constraints imposed by market 
economy. The full distribution uses add-on tools with various standard open-source licenses: LGPL, GPL, etc. 
Web address is \url{isabelle.in.tum.de}.
Isabelle unique features is a huge integrated environment for interactive and automated theorem proving. It is like a word-processor 
for formal logic, with specifications and proofs. Main user community is the people interested in formal 
logic and formalized mathematics and people doing proofs about software and hardware. Isabelle is in fact a multiplicity of ITP and ATP systems. 
Type of the system with respect to reasoning is mathematical, type with respect to logic is mostly HOL, but users can also do something else if 
they really want to. Its type with respect to
truth value is  mostly classical logic/Boolean. Programming paradigm is purely functional (ML and Scala) and UI is a full-scale IDE. 
It supports multiple operating systems such as: Linux, Windows, Mac and is available both for 32-bit and 
64-bit architectures. For scalability, it provides support for classic shared-memory workstations with many cores. 
Isabelle is highly modular, to the extent that it is hard to tell where it starts and ends and what is actually its true structure. 
It provides code generation facility for SML, OCaml, Haskell and Scala. Isabelle has a small proof kernel, according to the classic 
``LCF approach'', but with many add-ons and reforms over the decades. It is based on $\lambda$-calculus with simple types 
and natural deduction. Moreover, it supports inductive recursion and has very powerful derived principles for inductive sets, predicates, 
primitive and general recursive functions.
\\\\
{\bf Thierry Lecomte (Atelier B)}: Thierry Lecomte works as director at ClearSy organization. Under his supervision, ``Atelier B'' was developed. Atelier B 
implements the B method \cite{Abrial}  and offers a theorem prover. It was released first time in 1994. Latest version is
4.5.1 and updated last time in May 2018.  
Atelier B is an interactive rule based theorem prover
plus interactive and dedicated tableau method. Logic of the system is classical FOL and truth value is 
traditional Boolean. Calculus type is deductive and supports ZF set theory. Programming paradigm is imperative and 
programming language is similar to Prolog. Web address is \url{clearsy.com/en/our-tools/atelier-b}.
UI of Atelier B is graphical-based and it operates on various operating systems. It also provides
support for the Linux based clusters. Its architecture is monolithic, where inheritance and library
support is not available. Atelier B (CASE tool) provides C and Ada code generation. 
It has no small proof kernel but has proof-objects for more then 130 axioms. Mathematical rules (transformation, rewriting, hypotheses generation) 
are added by users, but it is not programmable by users. Syntax is inspired from Haskell, ML, Java, C, C++ and Prolog languages. 
Infix/postfix/mixfix operators' support are available and also for Unicode, Binary and ASCII coding schemes. Native support for B 
language is the unique feature of the system. Rich tactic language is available for writing proof instead by hand and it does not support inductive 
recursion.
\\\\
{\bf David Crocker (Escher Verifier)}: 
David Crocker is serving at MISRA C++ working group. He developed ``Escher'' verifier \cite{Cartlon}. Its latest version is 6.10.02 and 
was updated last time in 2015. It is an industrial product and license type of the system is commercial. Web address is 
\url{eschertech.com/products/ecv.php}. One of the unique feature of Escher Verifier is that sometimes it 
suggests missing preconditions/assertions/invariants, etc. in the model 
or software being verified when a proof is not found. Main user community is the
defense industry. It falls in ATP category. Reasoning type of the system is mathematical, logical type is a combination of FOL, SOL 
and some HOL. Type with respect to the truth value is mostly binary but triadic where necessary. Calculus of the system is
deductive and programming paradigm is mostly functional, but imperative in speed-critical parts. Programming language is
C++. There is no direct interface for the theorem prover. However, GUI is available for the verification tools that uses it. It runs on Windows
and Linux operating systems. System scalability is limited to a single thread and architecture is monolithic and standalone. 
It does not have small proof kernel and editor support. It is not extensible or programmable by its user. Moreover, it does not support
constructive logic. 
\\\\
{\bf Norman Megill (Metamath)}:
Norman Megill is the next respondent of the questionnaire. He is the originator of the ``Metamath'' \cite{Megill}. There are 34 other 
contributors who helped to extend the system. It was introduced first time in 1993. Latest version is 0.131 and was updated 
last time in June 2016. It is an independent development by Norman. Web page is \url{us.metamath.org}.
License type of the system is GPL. User FOL scheme is the unique feature of the system. 
It is an ITP and also used as a proof checker. Reasoning type of the system is mathematical, logical type is FOL. 
HOL is also possible but not developed yet. Truth value of the system is binary and deductive calculus
is used. There are 12 independent verifiers available in C, Java, C\#, Lua, Mathematica, Julia, Rust, Python, Haskell, C++, and 
JavaScript. Metamath supports both CLI, GUI and runs on almost all operating systems. 
Architecture of the system 
varies according to the environment. Metamath also displays comprehensive error message for the debug output and runtime assertion. 
Library of the system contains over 20000 theorems that covers results in logic, algebra, set and group theory, topology analysis, 
Hilbert spaces and quantum logic. It is a standalone system and no code generation facility is available. 
It is extensible and programmable by the user. 
Metamath has a small proof kernel. Human readability feature according to the syntax is unique as compared to others. Unification process is 
used for pattern matching. Argument handling is implicitly available and it is lightweight. Metamath supports inductive recursion and 
does not allow to write non-terminating programs. The system is easy to learn, but require experience with library and reasoning for advanced
proofs.
\\\\
{\bf Frank Pfenning (Twelf)};
Frank Pfenning works as professor at Computer Science Department, Carnegie Mellon University, USA and is the creator of ``Twelf''
\cite{Pfenning}. C. Sch\"{u}rman  also contributed to the system. It was released publicly in January 1999. Latest version is 1.7.1 and 
updated last time in January 2015. Website address is \url{twelf.org}. 
Twelf is an ITP and simplified BSD is the type of system license. Unique features of the system are meta-theorem proving for programming 
languages and logics. It is mainly developed for academia community. 
Type of the reasoning system with respect to logic is type theory and its type with respect to the truth value is intuitionistic. 
Twelf is built on deductive calculus and is developed in standard ML. 
UI of the system is CLI and runs on almost all operating systems. It may be scalable but it is not distributed 
or multi-threaded system. Twelf supports IDE and has its own libraries. System is extensible and programmable by users. 
It supports no tactic language and proofs are written by hand. Twelf has been used to formalize many different logics and programming
languages (examples are included with the distribution). 
\\\\
{\bf Ulf Norell (Agda)}:
Ulf Norell works as a principal research engineer at University of Gothenburg, Sweden. He developed ``Agda'' system \cite{Norell}. 
Latest version is  2.6.0.1, which was updated last time in May 2019. Web address is \url{wiki.portal.chalmers.se/agda/pmwiki.php}.  
License type of the system is BSD-like. Dependent types are the unique feature and academia is the main user community of the system. 
Popular research language is the main success story of the system.  It is an ITP and based on functional programming. System type  with respect 
to the logic is intuitionistic HOL, type with respect to the truth value is binary and is built on inductive calculus. 
It support constructive type theory. Haskell programming language is used for developing the system and UI of the system is graphic 
based. Agda supports and runs on all popular operating systems. It is scalable but not used for distributed or multi-threaded environment. 
It supports IDE, has its own proof kernel and library. It is extensible and programmable by users and has a tactic language for proof writing.
An important aspect of Agda is its dependence on Unicode. Its standard library is under constant development and 
includes many useful definitions and theorems about basic mathematical designs.
\\\\
{\bf Adam Naumowicz (Mizar)}: 
Adam Naumowicz provides his services to computer science institute at University of Bialystok, Poland. He gave information on ``Mizar'' 
\cite{Naumowicz}, 
which was publicly announced in November 1973. Latest version is 8.1.09 and updated last time in June 2019. 
Web address is \url{mizar.org}. Andrzej Trybulec is the founder and Mizar is developed at University of Bialystok. The system is free 
for any noncommercial purposes. User friendly input language based on natural language and a large library of formalized mathematics are 
the unique features. Mathematicians, computer scientists and students are the main user community. Mizar is an ITP  
and based on syllogism or mathematical statements. FOL with schemes (statements with free second-order variables) is the system type with respect 
to the logic and is based on binary truth value. It is based on deductive calculus and ZF set theoretic support is available. 
Declarative is the programming paradigm and object Pascal programming language is used for developing the system. Type of the 
interface is CLI and runs on almost all operating systems. It is scalable, but not suitable for distributed or multi-threaded 
environment. It supports IDE, has its own library and proof kernel. However, it is not extensible and no tactic 
language support is available for proof writing. Mizar Mathematical Library (MML) contains approximately 10,000 formal definitions 
and 52,000 lemmas and theorems.
\\\\
{\bf Michael Norrish (HOL)}: 
Michael Norrish has been working as principal research engineer at Australian National University. He talked about ``HOL'' theorem prover 
\cite{Norrish}. It was publicly released in January 1985. Latest version is Kananaskis-13 and was updated last time in August 2019. 
Web address is \url{hol-theorem-prover.org}. HOL was developed at Cambridge University. 
Four tools now comes in HOL family: HOL4 \cite{Slind}, HOL Light \cite{Harr}, ProofPower \cite{Arthan} and HOLZero \cite{Adams}. 
Other tools that come in HOL family are developed jointly by Cambridge University, 
Data61, CSIRO and Chalmers University of Technology. License type of HOL is BSD. It is an ITP based on syllogism 
or mathematical statements. HOL is the logical framework of the system and is based on binary truth value. 
System is based on deductive calculus and does not supports set theory directly, but it has a set-theoretic model. 
Programming paradigm is functional and developed in SML programming language. UI is command line. 
It supports and runs on all famous operating systems. It is scalable, but not suitable for distributed or multi-threaded environment. 
It does not support IDE, but has its own proof kernel and library. System is extensible and support tactic language for proof writing.
\\\\
{\bf Jonathan Sterling (RedPRL)}:
Jonathan Sterling is a graduate research assistant at School of Computer Science, Carnegie Mellon University and creator of 
``RedPRL'' \cite{Sterling}. Web address is \url{redprl.org}. 
MIT is the license type of the system. Unique features of RedPRL are higher dimensional types, support for strict equality and tactic scripts, 
refinement of dependent proofs and functional extensionality.
Main user community is homotopy type theory. 
It is an ITP and based on syllogism or mathematical statements. Type theory is the logical framework and is based 
on intuitionistic truth values. System is based on deductive calculus. Programming paradigm is functional and it is 
developed in standard ML. Visual studio code extension is the UI of the system. 
It runs on almost all major operating systems. RedPRL is not suitable for distributed or multi-thread environment, 
but has its own IDE. It has no library, but has its own proof kernel. RedPRL is extensible by the user and 
syntax of the system is inspired by Nurpl programming language \cite{Allen}. Tactic language support is also available for proof writing.
\\\\
{\bf Oleg Okhotnikov (Class \& Int Proof Checker)}:
Yuri Vtorushin and Oleg Okhotnikov implemented the ``Class and Int proof checker''. 
It was publicly announced first time in October 2007. Latest version of the system is Class 2.0 and Int 2.0 and was updated last time in 
November 2017. Web address is \url{class-int.narod.ru/}. Automated proof search for natural reasoning and support for iterative equalities
are the unique features of the system. System is mainly developed for students and teachers. Vtorushin and Okhotnikov uses Class and Int programs 
on seminars with students in courses ``Mathematical Logics and Algorithm Theory'', ``Artificial intelligence'', etc. It is an ATP based on syllogism 
or mathematical statements. It is based on FOL, supports binary truth value and built on deductive 
calculus. Axiomatic method set theoretic support is available. Programming paradigm is declarative and developed in C++. 
Moreover, it supports Windows operating system only and has a CLI. It is scalable and also supports distributed and multi-threaded 
environment. System supports IDE and has its own proof kernel. System is not extensible by the user and syntax is inspired by Mizar and SAD. 
Tactic language is also available for proof writing.
\\\\
{\bf Hans de Nivelle (GEO)}:
Hans de Nivelle from School of Science and Technology, Nazarbayev University, Kazakhstan developed the ``Geo'' \cite{Chou} prover.
It was released first time in August 2015 and latest version is Geo2016C. 
It is an ATP for FOL that is based on graph theory and 
supports partial classical logic (PCL) with 3-valued logic as a truth value. Calculus of the system is based
on geometric resolution and is
developed in C++. It supports CLI and only runs on Mac. 
The system takes geometric formulas and FOL formulas as input, where FOL formulas are changed to geometric formulas. 
During proof search, it looks for a geometric formulas model through backtracking.
Main success story is its existence in the current scenario.
License type is GNU GPL, Version 3. System is scalable, but not designed for distributed or multi-threaded environment and it
has no IDE. Library support is not available but proof kernel is owned by the system. Syntax of the system is inspired by
TPTP-language and it is not extensible by the user with no tactic support. Web page of the system is 
\url{http://www.ii.uni.wroc.pl/~nivelle/software/geo_III/CASC.html}.
\\\\
{\bf Hugo Herbelin (Coq)}: Hugo Heberlin is working as researcher at INRIA, France. He talked about the ``Coq'' system \cite{Bertot}. It was first released in May 1989. 
Latest version is 8.10.1 and last time updated in October 2019. 
Web address of the system is \url{coq.inria.fr}. It is developed by INRIA and academic partners. LPGL 2.1 is the license type. 
Unique features of the system are 
expressive logic and programming language, program extraction, elaborated certification 
language, proof techniques and a tactic language that allows users to define proof methods.
Teaching, formalization of mathematics and certified programming are main user community of the system. 
The specification language of Coq is called Gallina (based on Calculus of Inductive Constructions), 
which allows its users to write their specification by developing theories.
Coq follows the Curry-Howard isomorphism \cite{CHI} and uses Calculus of Inductive Constructions language \cite{CH88} to formalize programs, properties and proofs. 
Curry-Howard isomorphism provides a direct relation between programing and proving and says that proofs in a given subset of 
mathematics are exactly programs from a particular language. It means that one can use a programming paradigm to encode propositions and their proofs.
Coq is an ITP and supports various decision or semi-decision procedures produced proof-terms checked valid by a kernel. 
Logical framework of the system is based on HOL, $\lambda$-calculus and is built on both inductive as well deductive calculus. 
Different ways to represent sets are available in the system. Programming paradigm of the system is functional and developed in OCaml programming 
language. System supports both graphical as well as CLI and run on almost all operating systems. 
System is scalable, but not designed for distributed or multi-threaded environment. System has IDE, library support is available and 
proof kernel is also owned by the system. Syntax of the system is inspired by ML language and is extensible by the users.
\\\\
{\bf Clark Barrett (CVC4)}: Clark Barret is the last respondent of the questionnaire and provided information about CVC4 \cite{Barrett11}, developed at Stanford University and University of Iowa. 
CVC4 is an ATP for SMT problems and was released first time in December 2014. Latest version is 1.7 and last time updated in April 2019. 
Web address is \url{http://cvc4.cs.stanford.edu} and license type is BSD 3-clause.  
CVC4 is based on DPLL(T) calculus \cite{BarrettNOT06} and its type with respect to reasoning is mathematical and is based on standard many-sorted FOL, 
with limited support for HOL. It also supports finite sets.
Main user communities of the system are people that are interested in program analysis.
Programming paradigm is logical and is developed in C++. CVC4 supports CLI, offers API's for C, C++, Java, Python and runs on Mac and Windows. 
Support for finite sets is also available. Moreover, system is modular, does not provide any support for distributed computing and offers limited support 
for multi-threading.  
CVC4 offers solvers for separation logic, sets and relations, where models assign every formulas either {\em true} or {\em false}.  
Debug output and run time assertions support is available, whereas code generation support is not available. 
Input language to CVC4 is SMT-Lib, which is inspired by LISP. It also support CVC input language, which is more human-readable than SMT-LIB.
Moreover, limited support is available for inductive reasoning. CVC4 is used as the main engine in Altran SPARK toolset and at GE and Google. CVC4 comes first in 
various divisions of Satisfiability Modulo Theories (SMT-COMP), CASC and SyGuS (Syntax-Guided Synthesis) competitions.

The summary of the main characteristics of theorem provers for which we received answers from experts/developers is listed in Table 
\ref{modeltools5}. The answers for PVS is provided by authors of this paper as they have done some work in PVS in the past.

\begin{table}[!ht]
 \caption{Main characteristics of 16 theorem provers}\vspace{-3mm}
 \centering 
\rotatebox{90}{
\begin{tabular}{|p{3cm}|p{0.8cm}|p{0.8cm}|p{0.8cm}|p{0.8cm}|p{0.8cm}|p{0.8cm}|p{0.8cm}|p{0.8cm}|p{0.8cm}|p{0.8cm}|p{0.8cm}|p{0.9cm}|p{0.8cm}|p{0.8cm}|p{0.8cm}|p{0.8cm}|}
\hline
\textbf{Characteristics} & \tiny{\textbf{ACL2}} & \tiny{\textbf{E}} & \tiny{\textbf{Isabelle}} &
  \tiny{\textbf{Atelier B}} & \tiny{\textbf{Escher}} & \tiny{\textbf{Metamath}} & \tiny{\textbf{Twelf}} & \tiny{\textbf{Agda}} & 
  \tiny{\textbf{Mizar}} & \tiny{\textbf{HOL}} & \tiny{\textbf{RedPRL}} & \tiny{\textbf{Class \& Int}} & \tiny{\textbf{Geo}} & 
  \tiny{\textbf{Coq}} & \tiny{\textbf{PVS}} & \tiny{\textbf{CVC4}}\\ \hline
  System Type & \tiny{TP} & \tiny{TP} & \tiny{TP} & \tiny{TP} & \tiny{TP} &  \tiny{TP} & \tiny{TP} & \tiny{TP} & \tiny{TP} & \tiny{TP} & \tiny{TP} & \tiny{TP} & \tiny{TP} & \tiny{TP} & \tiny{TP} & \tiny{TP4SMT}\\\hline
  Theorem Prover Category & \tiny{ITP}  & \tiny{ATP} & \tiny{ATP+ITP} & \tiny{ITP} & \tiny{ATP} & \tiny{ITP} & \tiny{ITP} & \tiny{ITP} & \tiny{ITP} & \tiny{ITP} & \tiny{ITP} & \tiny{ATP} & \tiny{ATP} & \tiny{ITP} & \tiny{ITP}& \tiny{ATP} \\   \hline
  System Based on & \tiny{Syllogism} & \tiny{Syllogism} & \tiny{Syllogism} & \tiny{Syllogism} & \tiny{Syllogism} & \tiny{Syllogism} & \tiny{LF} & \tiny{FP} & \tiny{Syllogism} & \tiny{Syllogism} & \tiny{Syllogism} & \tiny{Syllogism} & \tiny{GT} & \tiny{DP} & \tiny{Syllogism} & \tiny{DPLL}\\\hline
  Logic Used & \tiny{FOL} & \tiny{FOL} & \tiny{HOL} & \tiny{FOL} & \tiny{FOL+HOL} & \tiny{FOL+HOL} & \tiny{TT} & \tiny{HOL} & \tiny{FOL} & \tiny{HOL} & \tiny{TT} & \tiny{FOL} & \tiny{PCL} & \tiny{HOTT} & \tiny{HOL} & \tiny{FOL}\\   \hline
  System's Truth Value & \tiny{Binary} & \tiny{Binary} & \tiny{Binary} & \tiny{Binary} & \tiny{Bin+Tri} & \tiny{Binary}  & \tiny{Intuition} & \tiny{Binary} & \tiny{Binary} & \tiny{Binary}
   & \tiny{Intuition} & \tiny{Binary} & \tiny{3-value} & \tiny{Binary} & \tiny{Binary} & \tiny{Binary}\\   \hline
  Calculus & \tiny{Inductive} & \tiny{Deductive} & \tiny{Ded+Indu} & \tiny{Deductive} & \tiny{Deductive} & \tiny{Deductive} & \tiny{Deductive} & \tiny{Inductive} & \tiny{Deductive} & \tiny{Deductive} & \tiny{Deductive} & \tiny{Deductive} & \tiny{Deductive} & \tiny{Ded+Indu} & \tiny{Deductive} & \tiny{Deductive}\\\hline
  Set Theoretic Support & \tiny{No} & \tiny{Yes} & \tiny{Yes} & \tiny{Yes} & \tiny{No} & \tiny{Yes}& \tiny{Yes} & \tiny{Yes} & \tiny{Yes} & \tiny{No} & \tiny{No} & \tiny{Yes} & \tiny{No} & \tiny{Yes} & \tiny{Yes}  & \tiny{Yes}\\   \hline
  Programming Paradigm & \tiny{Func} & \tiny{Impe} & \tiny{Func} & \tiny{Impe} & \tiny{Func+Imp} & \tiny{Func} & \tiny{LP} & \tiny{Func} & \tiny{Decl} & \tiny{Func} & \tiny{Func} & \tiny{Decl} & \tiny{Decl} & \tiny{Func} & \tiny{Func+OO} & \tiny{Logical}\\   \hline
  System Architecture & \tiny{Modular} & \tiny{Mod+Mono} & \tiny{Modular} & \tiny{Monolithic} & \tiny{Monolithic} & \tiny{Mod+Mono} & \tiny{Monolithic} & \tiny{Modular} & \tiny{Modular} & \tiny{Modular} & \tiny{Modular} & \tiny{Modular} & \tiny{Monolithic} & \tiny{Modular} & \tiny{Modular} & \tiny{Modular}\\   \hline
  Programming Language & \tiny{ACL2} & \tiny{C} & \tiny{ML+Scala} & \tiny{Prolog} & \tiny{C++} & \tiny{MM} & \tiny{SML} & \tiny{Haskell} & \tiny{Pascal} & \tiny{SML} & \tiny{SML} & \tiny{C++} & \tiny{C++} & \tiny{OCaml} & \tiny{C Lisp} & \tiny{C++}\\   \hline
  User Interface & \tiny{CLI} & \tiny{CLI} & \tiny{GUI} & \tiny{GUI} & \tiny{CLI+GUI} & \tiny{CLI+GUI} & \tiny{CLI} & \tiny{GUI} & \tiny{CLI} & \tiny{CLI} & \tiny{GUI} & \tiny{CLI} & \tiny{CLI} & \tiny{CLI+GUI} & \tiny{GUI} & \tiny{CLI} \\   \hline
  Platform Support & \tiny{Cross} & \tiny{Cross} & \tiny{Cross} & \tiny{Cross} & \tiny{Win+Linux} & \tiny{Cross} & \tiny{Cross} & \tiny{Cross} & \tiny{Cross} & \tiny{Cross} & \tiny{Cross} & \tiny{Windows} & \tiny{Mac} & \tiny{Cross} & \tiny{Mac+Linux} & \tiny{Mac+Win}\\   \hline
  Scalability & \tiny{Yes} & \tiny{Yes} & \tiny{Yes} & \tiny{Yes} & \tiny{No} & \tiny{Yes} & \tiny{Yes} & \tiny{Yes} & \tiny{Yes} & \tiny{Yes} & \tiny{No} & \tiny{Yes} & \tiny{Yes} & \tiny{Yes} & \tiny{Yes} & \tiny{No}\\   \hline
  Multi-threaded & \tiny{Yes} & \tiny{No} & \tiny{Yes} & \tiny{No} & \tiny{No} & \tiny{Yes} & \tiny{No} & \tiny{No} & \tiny{No} & \tiny{No} & \tiny{No} & \tiny{Yes} & \tiny{No} & \tiny{No} & \tiny{Yes}& \tiny{Yes}\\   \hline
  IDE & \tiny{Yes} & \tiny{Yes} & \tiny{Yes} & \tiny{Yes} & \tiny{No} & \tiny{Yes} & \tiny{Yes} & \tiny{Yes} & \tiny{Yes} & \tiny{No} & \tiny{Yes} & \tiny{Yes} & \tiny{No} & \tiny{Yes} & \tiny{Yes} & \tiny{No}\\   \hline
  Library Support & \tiny{Yes} & \tiny{Yes} & \tiny{Yes} & \tiny{No} & \tiny{No} & \tiny{Yes} & \tiny{Yes} & \tiny{Yes} & \tiny{Yes} & \tiny{Yes} & \tiny{Yes} & \tiny{Yes} & \tiny{No} & \tiny{Yes} & \tiny{Yes} & \tiny{Yes}\\   \hline
  Programmability & \tiny{Yes} & \tiny{No} & \tiny{Yes} & \tiny{No} & \tiny{No} & \tiny{Yes} & \tiny{Yes} & \tiny{Yes} & \tiny{No} & \tiny{Yes} & \tiny{Yes} & \tiny{No} & \tiny{No} & \tiny{No} & \tiny{Yes} & \tiny{No}\\   \hline
  Tactic Language Support & \tiny{Yes} & \tiny{No} & \tiny{Yes} & \tiny{Yes} & \tiny{No} & \tiny{Yes} & \tiny{No} & \tiny{Yes} & \tiny{No} & \tiny{Yes} & \tiny{Yes} & \tiny{Yes} & \tiny{No} & \tiny{Yes} & \tiny{Yes} & \tiny{No}\\   \hline
   \end{tabular}

\label{modeltools5}
}
\end{table}

We also developed a layout for the survey questionnaire, which is listed in Table \ref{slrl}. Mnemonics codes are used to represent headlines. 
These abbreviations are: CLang = Computational Language, 1st Rel = First Release,
Ind/Uni/Inde = Industry/University/Independent, Prog.P = Programming Paradigm, LV = Latest Version, 
LT = License Type, UI = User Interface, OS = Operating System, Lib = Library, CG = Code Generation, Ed = Editor, Ext = Extendable, I/O = 
Input/Output,
TType = Tool Type, CLogic = Computational Logic, TV = Truth Value, ST = Set Theory, App.Areas = Application Areas and Eval = Evaluation. 
We filled the layout for 27 more theorem provers. We collect the data from various resources such as electronic databases, research articles 
and dissertations. Complete detail for each system is listed in Appendix  \ref{A1}.

\begin{table}[!ht]
 \caption{Systematic literature review design}\vspace{-3mm} 
 \centering 

\begin{tabular}{|p{0.01\textwidth}|p{0.135\textwidth}|p{2cm}|p{2cm}|}\hline
\multicolumn{4}{|c|}{Theorem provers} \\
\hline
\parbox[t]{2mm}{\multirow{4}{*}{\rotatebox[origin=c]{90}{\textbf{General}}}} & Name & & \\
 & Contributor & & \\
 & 1st Rel & &  \\
 & Ind/Uni/Ind & & \\ \hline
\parbox[t]{2mm}{\multirow{11}{*}{\rotatebox[origin=c]{90}{\textbf{Implementation}}}} & CLang &  &  \\
 & Prog.P & &  \\
 & LV &  &  \\
 & LT &  &  \\
 & UI &  & \\
 & OS &  &  \\
 & Lib & &  \\
 & CG &  & \\
 & Ed &  & \\
 & Ext &  &  \\ \hline
\parbox[t]{2mm}{\multirow{5}{*}{\rotatebox[origin=c]{90}{\textbf{Logico-Math}}}} & TType & &  \\
 & CLogic &  & \\
 & TV &  & \\
 & ST &  & \\
 & Calculus & & \\
 & ProofKernel &  & \\ \hline
\parbox[t]{2mm}{\multirow{3}{*}{\rotatebox[origin=c]{90}{\textbf{Others}}}} & App. Areas & &  \\
 & Eval &  &
  \\
& Unique Features & &
 \\
\hline
\end{tabular}
\label{slrl}
\end{table}

\section{Comparison} \label{s5}
In this section, we first listed those scientists that contributed most in the formalization of mathematical theorems. 
We also present those theorem provers in which most of the theorems are proved. Furthermore, the top systems 
(from 1996 till 2019) in CADE ATP system competition are described. Finally, we showed the comparison of more than 40 provers for 
parameters mentioned in Section \ref{s1}. 

\subsection{Top Scientists and Theorem Provers}
\begin{figure}[!ht]
\centering
  \includegraphics[width=9cm]{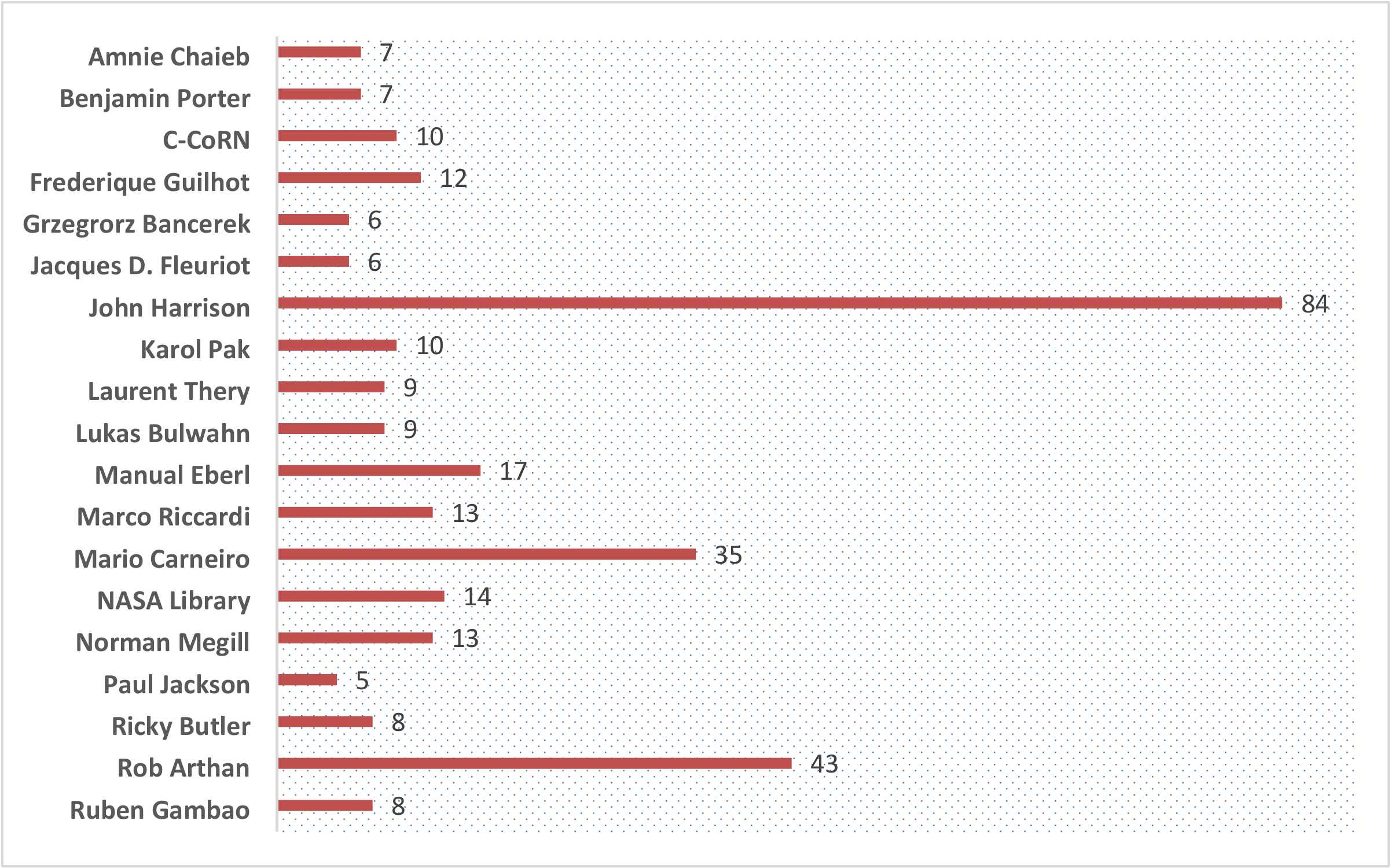}
  \caption{Theorems proved by scientists in ITPs}
  \label{phpt}
\end{figure}

Efficiency and power of theorem provers are generally evaluated on the number of theorems they proved from top hundred theorems 
list (available at: \url{http://www.cs.ru.nl/~freek/100/}). People who contributed most in verifying these theorems are presented 
in Figure \ref{phpt}. John Harrison currently working at Intel proved 84 theorems and he used HOL (particularly HOL Light) and Isabelle. Rob Arthan 
(second in the list) also used 
family of HOL theorem provers for theorem proofs. Theorem provers which proved most of the theorems from top hundred theorems
list are presented in the order: 
\begin{center}
HOL Light (86) $\rightarrow$ Isabelle (81) $\rightarrow$ Metamath (71) $\rightarrow$ Coq (69) $\rightarrow$ Mizar (69) $\rightarrow$ ProofPower (43)
$\rightarrow$ Nqhtm/ACL2 (18) $\rightarrow$ PVS (16) $\rightarrow$ NuPRL/MetaPRL (8) 
\end{center}

HOL Light's performance is outstanding and it is at the top by formalizing 
and proving 86 theorems. Isabelle, another powerful tool is at the second number in the list. Metamath, Coq, Mizar and ProofPower are also 
the computationally strong tools and play a vital role in the formalization of top hundred theorems. 
Two systems in HOL family (HOL Light and ProofPower) are included in the list.

\subsection{Best FOL Theorem Provers at CASC}
\begin{figure}[!ht]
  \centering
  \includegraphics[width=9cm]{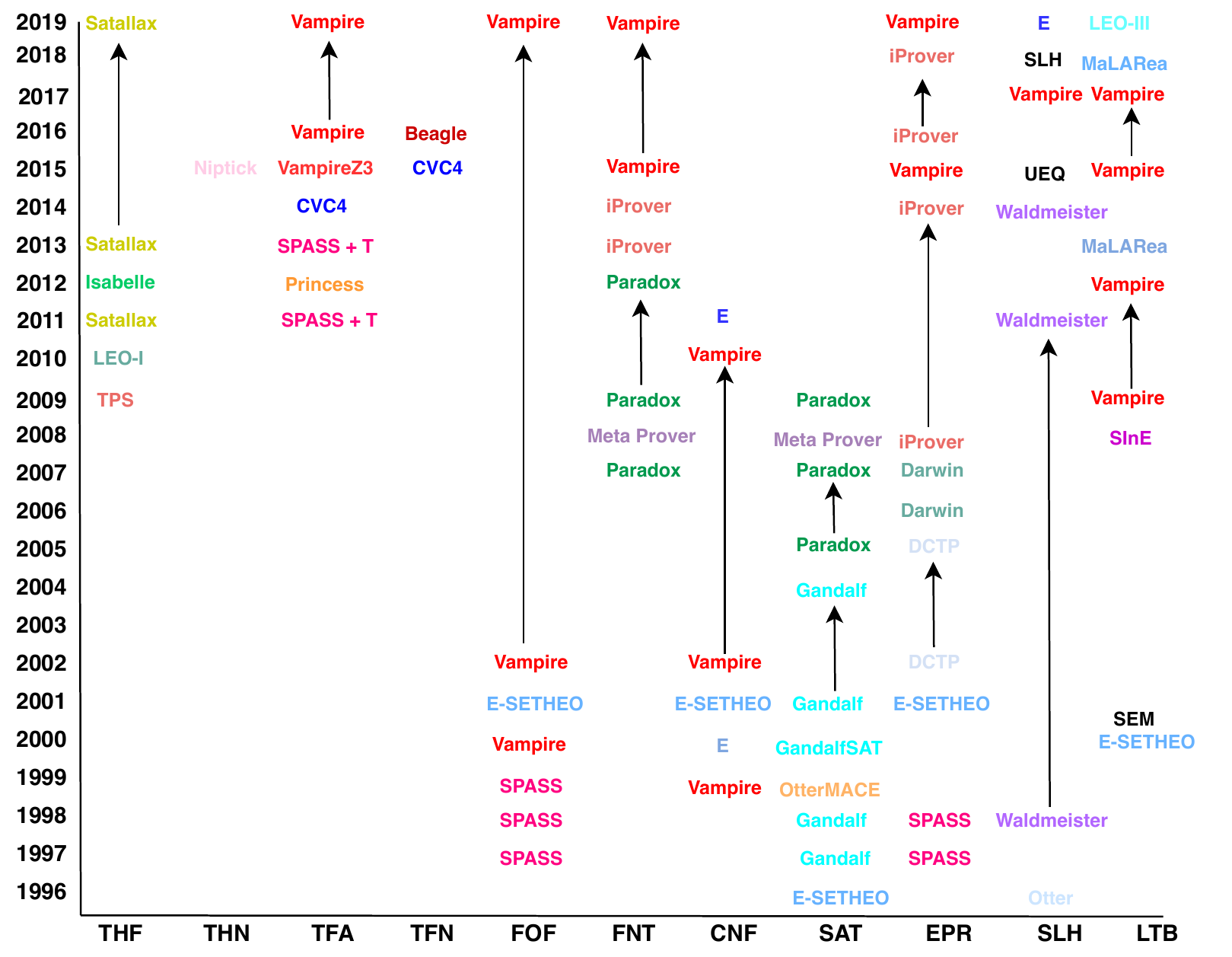}
   \caption{Top ATPs at CASC}\label{SC}
\end{figure}

Each year, FOL based ATPs performances are checked in the CADE ATP System Competition (CASC). 
This competition was started in 1996. CASC consists of various divisions and these divisions are categorized based on the type of 
their problems and the characteristic of systems. There are two major divisions. First one is the competition division which ranks the reasoning
system, second one is the demonstration division which enables the system to demonstrate its potential without ranking. These divisions are 
further divided into subdivision on the basis of problem categories. Competition division is an open platform for automated reasoning systems 
that meet the requirements of this division. System selected for the competition division tries to attempt all the problems of this division. 
Subdivisions of this division are: THF, THN, TFA, TFN, FOF, FNT, CNF, SAT, EPR, SLH (changed to UEQ in 2015 and back to SLH in 2018) and LTB. More details on 
subdivisions can be found in \cite{Sutcliffe16}. These divisions are presented on horizontal 
axis in Figure \ref{SC}, while vertical axis represents the competition year. Some tools are specified to only one division, 
while some are tested on different problem divisions which show excellent results. Figure \ref{SC} has sketched the overall results 
for each division. Arrows in the figure show the continuous winners in a particular division. For example, Vampire \cite{Riazanov} in 
the FOF division is performing best from 2002 to 2019 and Satallax \cite{Brown} is coming first in the THF division from last seven years.
Vampire system topped the TFA, FOF, FNT and EPR divisions respectively. Similarly, iProver\cite{Korovin08} dominates the EPR division from 2008 to 2014 and 2016 to 2018.
These provers perform best at CASC due to the following reasons:
\begin{enumerate}
 \item Sound theoretical foundations, \item Thorough tuning and testing, \item Huge implementation efforts, and \item 
Understanding of how to optimize for the competition.
\end{enumerate}

\subsection{Provers Comparative Analysis Results}
This subsection presents the comparative analysis of theorem provers for various parameters, that includes: theorem prover category, 
mechanical reasoning system type, logical framework, truth value, calculus of the system, set theoretic support,
programming paradigm and programming language of the system, UI, system scalability, distributed/multi-threaded,
IDE and library support. Results for each parameter is presented next. 

\subsubsection{Theorem Prover Category}
Figure \ref{TheoremProverCategory} shows the category of theorem provers such as ATPs, ITPs, geometric theorem provers, decision procedures
and theory generators, etc. Major portion is shared by ATPs (44\%) and ITPs (31\%). Systems working as both ATP and ITP take 11\%. 
Whereas, systems that work as either theory generator, as ATP and model generator and as automated geometric theorem prover
share 6\%. Systems that work as decision procedures and ATP for SMT problems take 8\%.

\subsubsection{Mechanical Reasoning System Type}
Figure \ref{Mechanical Reasoning System type} represents the theorem provers grounding either these are based on syllogism, mathematical 
statements or various logic theories. 54.5\% of the systems are based on syllogism or mathematical statements. While other types take 9.1\%  
each.
\subsubsection{Logical Framework}
Figure \ref{Logical Framework Mechanical Reasoning System} shows the logical framework of theorem provers. Most systems (59\%) are FOL 
based systems. 16\% of the systems are based on HOL. Systems that are 
based on graph theory, dynamic modal logic, FOL/HOL, pure classical logic, equational logic, type theory and higher order type theory 
share collectively 25\%.
\subsubsection{Truth Value and System Calculus}
Figure \ref{Truth Value of Mechanical Reasoning System} shows 86\% of theorem provers are based on Boolean logic or binary logic, while 7\% of the 
systems are based on intuitionistic logic. Systems that result in triadic (3-value) share 2\% and both binary and 
triadic value systems take 5\%.

Figure \ref{Calculus of the System} shows 52\% of the systems are based on deductive calculus while 10\% are based on inductive calculus. 
Theorem provers which are based on both inductive and deductive calculus take 5\%. Systems based on euclidean and differential geometry, 
first order predicate calculus, fixed point co-induction, $\lambda$-calculus, sequent calculus, tableau calculus, instantiations calculus, 
hyper tableau calculus and typed $\lambda$-calculus are respectively 3\%, 3\%, 3\%, 5\%, 5\%, 8\%, 2\%, 2\% and 2\%.


\subsubsection{Set Theoretic Support}
Figure \ref{Set Theoretic Support} shows that only 30\% of theorem provers provide set theoretic support, while 56\% do
not support set theory. Systems that support  Horn theory, swinging type theory and ZF set theory take 2\% each. Systems that 
supports Quine's and B-Method set theory also take 3\% each.
\subsubsection{Programming Paradigm and Programming Language}
\begin{figure*}[!htp]
 \begin{subfigure}{.5\textwidth}
  \centering
  \includegraphics[width=.60\linewidth]{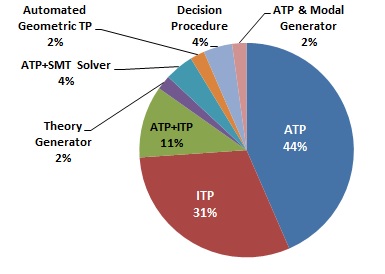}
  \caption{Theorem provers category}
  \label{TheoremProverCategory}
\end{subfigure}%
\begin{subfigure}{.5\textwidth}
  \centering
  \includegraphics[width=.81\linewidth]{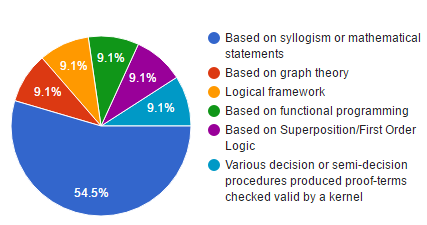}
  \caption{Mechanical reasoning system type}\label{Mechanical Reasoning System type}
\end{subfigure}
  \centering
  \begin{subfigure}{.5\textwidth}
  \centering
  \includegraphics[width=.66\linewidth]{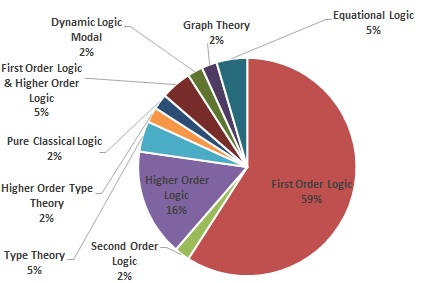}
  \caption{Logical framework of systems}\label{Logical Framework Mechanical Reasoning System}
\end{subfigure}%
\begin{subfigure}{.5\textwidth}
  \centering
  \includegraphics[width=.66\linewidth]{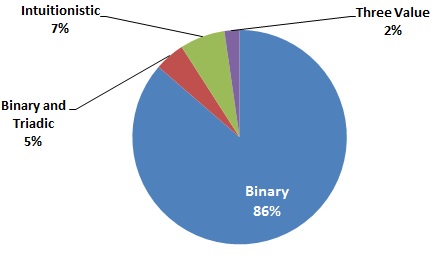}
  \caption{Truth value of the systems}\label{Truth Value of Mechanical Reasoning System}
\end{subfigure}
\begin{subfigure}{.5\textwidth}
  \centering
  \includegraphics[width=.66\linewidth]{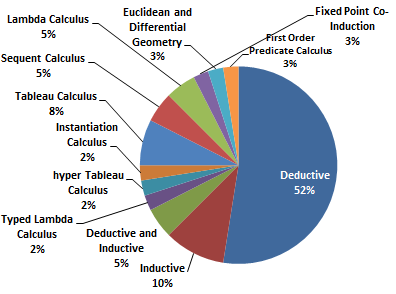}
  \caption{Calculus of the systems}\label{Calculus of the System}
\end{subfigure}%
\begin{subfigure}{.5\textwidth}
  \centering
  \includegraphics[width=.66\linewidth]{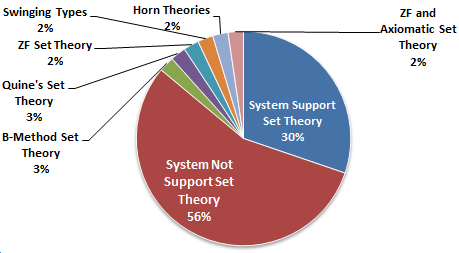}
   \caption{Set theoretic support}\label{Set Theoretic Support}
\end{subfigure}
\begin{subfigure}{.5\textwidth}
  \centering
  \includegraphics[width=.65\linewidth]{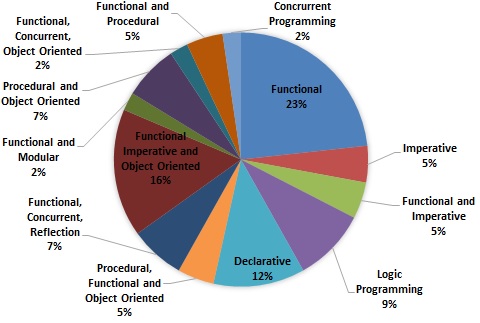}
  \caption{System programming paradigm}\label{Programming Paradigm of the system}
\end{subfigure}%
\begin{subfigure}{.5\textwidth}
  \centering
  \includegraphics[width=.65\linewidth]{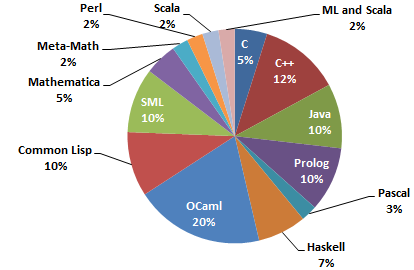}
   \caption{System programming language}\label{Programming Language of the system}
\end{subfigure}
\begin{subfigure}{.5\textwidth}
  \centering
  \includegraphics[width=.85\linewidth]{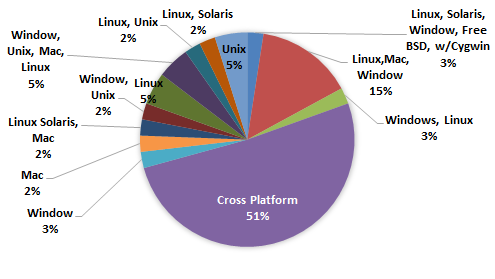}
  \caption{Operating system}\label{Operating System}
\end{subfigure}
  \centering
\caption{Theorem provers comparisons}
\end{figure*}

Figure \ref{Programming Paradigm of the system} presents the programming paradigm of the systems such as functional, imperative and declarative, etc.
Programming paradigms of 23\% theorem provers are functional. While systems having functional, imperative and object oriented paradigms take 
16\%. Systems that belong to logic programming paradigm take 9\%. Provers that belong to both procedural and object oriented paradigm are 
7\% and 12\% systems belong to the declarative paradigm. 2\% of the systems belong to functional, concurrent, 
and object oriented programming paradigm. Systems that belong to functional and imperative paradigm take 5\%. 
Theorem provers which come under the functional and procedural 
paradigm take 5\%. Systems that only belong to concurrent programming paradigm take 2\%. Systems belonging to both functional and modular 
paradigm take 2\%.

Figure \ref{Programming Language of the system} presents the overview of programming languages which are used to develop theorem provers:  
Ocaml (20\%), C/C++ (17\%), Common Lisp (10\%), Java (10\%), Prolog (10\%), SML (10\%), Haskell (7\%), Mathematica (5\%), Pascal (3\%),
Metamath (2\%), Perl (2\%), Scala (2\%), ML and Scala (2\%).
\subsubsection{User Interface and Operating System}
Theorem provers are mostly available with CLI (54\%)  only, while 30\% percent of the
systems are available with GUI only and 16\%  of the systems provide both CLI and GUI.

On the other hand, 51\% theorem provers run on cross platform as shown in Figure \ref{Operating System}. 15 \% of the systems
support Linux, Mac and Windows operating systems. 3\% of the systems run on all Unixoids-based operating systems. Systems that run only
on Windows take 3\%, Unix (5\%), Linux (5\%), and Mac (2\%). Systems that support Unix as well as Linux are 2\%. Systems that 
run on Linux, Unix, Windows and Mac are 5\%. Systems supporting Linux, Solaris and Mac take 2\%. Systems that only support Linux and Solaris
take 2\%. 

\subsubsection{Distributed/Multi-threaded, IDE and Library Support}
Only 17\% (8 out of 43) of theorem provers support distributed or multi-threaded environment,
while 83\% of the systems does not support such environments. Further, our results showed that all of the systems have the ability of scalability
according to future needs.
On the other hand, 65\% (30 out of 43) of the systems support IDE.
Furthermore, 56\% (26 out of 43) of the systems have their own libraries while 44\% of the systems does not have their own libraries.

\subsection{The de Bruijn Criterion}
According to the de Bruijn criterion ``the correctness of the mathematics in the system should be guaranteed by a small checker'' \cite{Baren}.
This means that a system has a `proof kernel' (also called proof checker) that is used to filter all the mathematics. 
Table \ref{spk} shows whether 15 theorem provers for which we received answers from experts/developers have small proof kernels or not 
+ stands for yes and - for no). Proof kernel for other 27 theorem provers are shown in Appendix \ref{A1}, and majority of them have no proof kernel. 
Whereas, HOL Light has extremely small proof kernel containng only several hundred lines of OCaml. 

\begin{table}[!ht]
 \centering
 \caption{de Bruijn Criterion}\label{spk}\vspace*{-2mm}
  \begin{tabular}{|c|c|c|}
    \hline
    {\bf System} & {\bf De Bruijn Criterion} & {\bf Proof-object} \\ \hline
     ACL2 & - & - \\\hline
     E & - & + \\\hline
     Isabelle & + & x \\\hline
     Atelier B & - & - \\\hline
     Escher & - & - \\\hline
     Metamath & + & + \\\hline
     Twelf & - & - \\\hline
     Agda & + & + \\\hline
     Mizar & - & - \\\hline
     HOL & + & x \\\hline
     RedPRL/Nuprl & + & x \\\hline
     Class \& Int & + & +\\\hline
     Geo & + & + \\\hline
     Coq & + & + \\\hline
     PVS  & - & - \\\hline
     \end{tabular}
\end{table}


In some ITPs (e.g., Coq and Agda), 
the proof kernel also checks the correctness of {\em proof-objects} that are generated by other tools included in the whole system. 
For an ITP with {\em proof-objects}, the proof-script for a statement (theorem or lemma) contain a list of tactics/strategies that are required to make the 
proof-assistant to verify the validity of the statement. The proof-script generates and stores
a term that is a proof that can be checked by a simple proof kernel. The reliability of the whole system depends on the
soundness of {\em proof-objects} and the proof kernel. 
Even if someone have doubts about the validity of certain statements or if some parts of the systems contain bugs, the {\em proof-object} for a given statement 
and the proof kernel can be used to locally verify the statement within the corresponding logical system.

HOL, Isabelle and Nuprl come in the class of ITPs that have a proof kernel but no {\em proof-objects}. In such systems, the 
proof-script are considered 
as {\em no-standard proof-object} (shown with x in Table \ref{spk}). They translate the proof-script into a {\em proof-object} that requires some system specific 
preprocessing. The trustworthiness of the translation is verified with the proof kernel. For ITPs with no proof kernel (e.g., PVS, Mizar and ACL2), 
there is no way (yet) that provides a {\em proof-object} with high reliability. One has to trust these systems for the correctness of statement accepted by the 
assistants. The advantage of these kind of systems generally is their larger automated deduction facilities and user-friendliness \cite{Barendregt01}.



\subsection{The Poincar{\'e} Principle and Automation}
For theorem provers, one of the important aspects is the automation of trivial tasks \cite{Wiedijk}. It means that a user 
is not required to explain all the details of the calculations to a theorem prover. A theorem prover satisfies the Poincar{\'e} principle 
(formulated by \cite{Barendregt12}) if it has the ability to automatically prove the correctness of calculations.  For example 3 + 4 = 7 holds 
by computation and it should not be justified with long chains of logical inferences.
Table \ref{sa} lists whether a prover satisfies Poincar{\'e} principle or not.

\begin{table}[!ht]
 \centering
 \caption{The Poincar{\'e} principle}\label{sa}\vspace*{-2mm}
  \begin{tabular}{|c|c|c|}
    \hline
    {\bf System} & {\bf Poincar{\'e} principle} & {\bf User automation} \\ \hline
     ACL2 & + & + \\\hline
     E & - & + \\\hline
     Isabelle & + & + \\\hline
     Atelier B & + & + \\\hline
     Escher & - & - \\\hline
     Metamath & - & - \\\hline
     Twelf & - & - \\\hline
     Agda & - & - \\\hline
     Mizar & - & - \\\hline
     HOL & + & + \\\hline
     RedPRL/Nuprl & + & - \\\hline
     Class \& Int & + & - \\\hline
     Geo & - & + \\\hline
     Coq & + & + \\\hline
     PVS  & + & + \\\hline
     \end{tabular}
\end{table}
  

Another important feature of a theorem prover is whether it enables its user to write programs that can solve proof
problems algorithmically. HOL, PVS, Coq and Isabelle offers such kind of user automation.
Strong built-in automation (proof tactics, decision and search procedures, and induction automation etc.)
is another important aspect of a theorem prover.
ACL2 and PVS has the powerful built-in automation. Then comes HOL, Isabelle and then Mizar and Coq.
The following order indicates the reliability of ITPs with most reliable on the left side and least reliable on the right side.
\begin{center}
Agda, Coq, Metamath, Nuprl, HOL, Isabelle, Twelf, Mizar, PVS, ACL2.
\end{center}

Agda is placed at first place as it only uses predicative logic. The middle places are occupied by Nuprl, HOL and Isabelle 
because of their {\em non-standard proof-objects}. Finally, the least reliable are those that do not work wit {\em proof-objects}. On the other hand, the order 
for internal automation in ITPs is opposite with ACL2 and PVS on the top.

\section{Strengths, Analysis, Applications and Future Directions}\label{SAFD}
In this section, we provide the details on the strengths, in-depth analysis of the differences and applications of theorem provers. 
Moreover, some future research directions based on recent work is also discussed. Some theorem provers such as Geo, Class $\&$ Int proof 
checker are omitted in this section and some other famous provers such as Nuprl, Vampire, Prover9/iProver and MaLARea are included. 

\subsection{ITPs}
ACL2 main strengths are: state-of-the-art prover heuristics, robust engineering and extensive hyperlinked documentation. 
Among all ITPs, ACl2 uses FOL instead of higher-order logics. Only two ITPs, Isabelle and ACl2 offer parallel proof checking facility. 
ACl2 also supports program extraction (also called code generation) by translating the specification in ACL2 language to Common Lisp. Theories can also be developed and
executed in ACL2 as it is built around a real programming language. Users can not construct inductive types, but powerful built-in induction scheme in ACl2 allows users 
to define their own recursive functions. The inference engine is based on the waterfall design of Nqthm prover \cite{PF18}.
ACL2 has been used extensively to verify hardware and software designs at AMD, Centaur, Oracle, Intel, IBM and to prove separation 
kernel properties at Rockwell Collins \cite{Hunt}. Moreover, ACL2 is also used successfully in processor modeling \cite{Flatau}, 
digital systems \cite{Kaufmann3}, programming languages \cite{Moore}, asynchronous circuits \cite{Chau17} and 
concurrency \cite{Moore2}. ACl2(ml) \cite{Heras} uses machine learning to facilitate users in the proof process.
In past, some work has been done in integrating SAT solvers into ACL2 \cite{PengG15, Swords}. 
However, it has generated new issues because of their support for a wide range of domains including real numbers and uninterpreted 
functions. The $x86isa$ library  \cite{Goel} in ACL2 offers a formal model for reasoning about $x86$ machine-code programs. 
Adding several features to current $x86isa$ library such as exceptions, interrupts handling and extending I/O capabilities will enable 
us to reason about real system codes. 

Atelier B offers a framework that automatically prove and review user added mathematical proof rules. 
Proof obligations in Atelier B contain traceability information that helps in locating modeling errors
and model editor allows the navigation of models and operations \cite{Lecomte17}. B-method allows one to develop many models of the same system 
with the refinement technique. However, one is required to explain the B model while proving theorems and the proof obligation generator may generates small 
proof obligations that need to be discharged.
Atelier B is used in the development of safety-critical systems \cite{Lecomte2} and communication protocol \cite{Julliand}.
Moreover, B-method is also used successfully in the development of safety-critical parts in 
two railway projects \cite{Abrial3, Badeau} and byte code verifier of the Java card \cite{C02}. 

Main strength of Metamath is that it uses the minimum possible framework that is required to express mathematics and their proofs. 
Unlike most ITPs, 
no assumption is made by the Metamath's verification engine about the underlying logic and the main verification algorithm is very simple (essentially nothing more
than substitution of variables expression, enhanced with checking for conflicts in distinct variables). 
Weaker logics such as quantum or intuitionistic can be handled in Metamath with different sets of axioms. 
Proofs in Metamath are generally very long, but the proofs are completely transparent and Metamath database contains over 30,000 human readable formal proofs.
In the proof development process, users prove a theorem/lemma interactively within the program, which is then written to the source by the program.
A definition is provided in \cite{Carneiro15} for models of Metamath style formal systems, which are demonstrated on propositional
calculus examples. 
From mathematical foundations, Hilbert space and quantum logic is developed in Metamath, which are used in the verification of some
new results in these fields. Metamath is used in the formalization of Dirichlet's theorem and Selberg's proof 
of the prime number theorem \cite{Carneiro}. 
An algorithm is presented in \cite{Carneiro16} that converts HOL Light proofs into Metamath.

Twelf strengths are representing deductive systems with side conditions and judgments with contexts. It offers an environment for experimenting with encodings 
and to verify their meta-theoretic properties. It also provides a module system for the organization of large developments. 
Twelf implements the {\em $\lambda$Prolog} \cite{FGHMNS88} and its logic is very close to the Edinburgh Logical Framework (LF) \cite{Pfenning}. 
Twelf specifications can be executed through a search procedure, which means that it  can also be used as a logical programming language.
Twelf is used in proving the safety of standard ML programming 
language \cite{Lee}, in typed assembly language system \cite{Crary}, in foundational Proof Carrying-Code system \cite{Appel},
in cut-elimination proofs for classical and intuitionistic logic \cite{Pfenning95}, for specifying and validating logic
morphisms \cite{Schurmann6} and construction of a safety policy for the IA-32 architecture \cite{Crary08}.

Main features of Agda is the interactive construction of programs and proofs with meta-variables and place-holders. 
Unlike other ITPs that work with {\em proof-scripts}, Agda acts as a structure editor, providing support for term construction. 
Users can edit the {\em proof-object} by focusing on a hole and executing one of the operations (tactics) that is applied to that hole. It is the only ITP that offers a 
functional programming language with dependent types.
Moreover, strictly positive inductive and inductive-recursive data types are supported in Agda.
Emacs interface for Agda allows the incremental development of programs \cite{Bove}.
Agda is used to formally verify railway interlocking systems \cite{Kanso}, web client application development \cite{Jeffrey},
fully certified merge sort \cite{Copello}, hardware description and verification \cite{pizaniflor}, 
formalizing Type-Logical Grammars \cite{Kokke}, Curry programs verification \cite{Antoy} and formalization of Valiant's Algorithm \cite{Philippe}.
In \cite{Foster}, automated theorem prover Waldmeister \cite{Hillenbrand} is integrated in Agda to facilitate the proof automation. 
Similarly, another tool is proposed in \cite{Lindblad} for automated theorem proving in Agda.

Mizar received popularity because of its huge repository of formalized mathematical knowledge, 
which has been used in developing AI/ATP methods to solve conjectures in mathematics \cite{Josef23a}. 
Over the years, the syntax of Mizar is improved, simplified and the Mizar language now contains
a rich set of logical quantifiers and connectives.
The evolution of Mizar in first 30 years is presented in \cite{Matuszewski5}. 
In Mizar, proofs are written in a declarative way and proofs are developed according to the rules of the Ja\`{s}kowski style of
natural deduction \cite{Jaskowski}. This characteristics influenced other systems to build similar kind of proof
layers on top of several other systems, such as the Mizar mode for HOL \cite{Harrison05}, the Isar language for Isabelle \cite{Wenzel02}, 
Mizar-Light for HOL-Light \cite{Wiedijk01} and the declarative proof language (DPL) for Coq \cite{Corbineau}. 
Mizar does not support the Poincar{\'e} principle, yet it has some automated deduction and a set of tactics that are very user-friendly. 
The main unique feature is that the {\em proof-script} is close to an ordinary proof in mathematics.
Apart from large MML, Mizar is used in the development of rigorous mathematics, in hardware/software verification and in mathematical education \cite{Naumowicz}.
Some recent work in Mizar includes the formalization of Pell's equation \cite{Acewicz}, Nominative Algorithmic Algebra \cite{Kornilowicz}
and formalization of bounded functions for cryptology \cite{Okazaki}. 
A suite of AI/ATP system is developed on Mizar library in \cite{Kaliszyk15} that contains approximately 58000 theorems. 
14 strongest methods that executed in parallel proved 40.6\% of the theorems.
Moreover, an independent certification mechanism is developed in Mizar that is based on Isabelle framework \cite{Kaliszyk17}. 
In \cite{Kaliszyk160}, Mizar environment is also emulated inside the Isabelle.

HOL was build upon the Cambridge LCF approach \cite{PA87} that is now referred to as HOL/88 with the purpose of hardware verification. It has influenced the 
development of other famous ITPs such as Isabelle/HOL,
HOL Zero, ProofPower and HOL Light. In HOL, the computations that involves recursion can become quite lengthy and complex when 
they are converted to {\em proof-objects}. Thus, the {\em proof-objects} are not stored, only the {\em proof-scripts}. 
This is the reason why {\em non-standard proof-objects} are used in HOL.
Users in HOL generally work inside the implementation language. As HOL is fully programmable, various other means of interacting 
with HOL have also been developed. HOL Light is the most widely used provers of this family and it is probably the only prover that represents the 
LCF approach in its purest form. Its logic is based on simple type theory with polymorphic type variable. The terms in HOL Light are of simply typed 
$\lambda-$calculus, 
with just two primitive types: {\em bool} (Booleans) and {\em ind} (individuals). HOL has been used extensively for formal verification projects in industry.
HOL provers are used widely in the formalization of mathematical theorems \cite{Avigad}, hardware design \cite{Harrison5, Hamid},
communication protocols verification \cite{Hasan2}, programs \cite{Harrison6} 
and control system analysis \cite{Rashid}. 
Similarly, HOL(y) Hammer offers machine learning based premise selection and automated reasoning both for HOL Light and HOL4 \cite{Kaliszyk15}.
Recent work in HOL includes the formalization of quaternions \cite{Gabrielli17}, linear analog circuits \cite{Taqdees17},
process algebra CCS \cite{Tian17a} and metric spaces \cite{Maggesi18}. A library for combinational circuits is developed in \cite{Shiraz18}.
Moreover, formalized Lambek calculus has been ported from Coq to HOL4 in \cite{Tian17} with some new theorems and improvements. 
A technique based on A* algorithm is presented in \cite{Gauthier17} to automate the selection process of 
tactics and tactic-sequences in HOL4.

Nuprl \cite{Constable}, which inspired the development of RedPRL, is a proof development system based on Computational
Type Theory. Nuprl has evolved significantly in years and now can handle those logics 
where inference rules can be declared in a sequent style. It follows the LCF approach and the type theory include less-common subtype, 
very-dependent function types and type constructors of quotient type. Type checking in Nuprl is undecidable as subtypes can be defined with arbitrary types. 
Whereas in PVS, an algorithm for type-checking automates all simpler type checking tasks. Moreover, the computation language is untyped and judgments are also not 
decidable because the Poincar{\'e} principle is assumed not only for {\em intensional equality} but also for {\em extensional equality}. Users can interact
with Nuprl only through structural editors where proofs can be edited and viewed as proof trees.
Nuprl is used in mathematics formalization and hundred of theorems are proved in the system \cite{ACEKL00}. 
Nuprl is also used in protocol verification \cite{BKVL01}, hardware and software specification and 
verification \cite{Aagaard, Lesser10}, reasoning about functional programs \cite{Howe}, design of reliable
networks \cite{Kreitz08}, and the development of demonstrably correct distributed systems \cite{BG05}.
Some work on the integration of Nuprl with other systems is done in the past. 
In \cite{ABCEK03}, PVS is integrated in Nuprl to enables users to access PVS from the Nuprl environment. 
A new semantics is provided in \cite{Howe6} to embed the logic of the HOL prover inside Nuprl.
Similarly, Nuprl's meta-theory is formalized in Coq \cite{Anand}, which is later used in the Nuprl proof for the validity of Brouwer's Bar 
Induction principle \cite{Rahli}.

Coq also follows the LCF approach and probably the most developed ITP after HOL Light. 
The logic used in Coq is very expressive that can define rich mathematical objects. 
Moreover, Coq has the ability to explicitly manipulate {\em proof-objects} from {\em proof-scripts}, 
which makes the integrity of the syetm dependent on the
correctness of the type-checking algorithm. As Coq is based on constructive foundations, it has two basic meta-types (also called sorts): {\em Prop} (as a type of 
logical propositions) and {\em Set} (as a type of other types (eg., Booleans, naturals, subsets, etc) \cite{Y18}.
This allows Coq to distinguish between terms that represent proofs and terms that represent programs.
A program extractor can also be used to synthesize and extract verified programs (in OCaml, Haskell or Scheme) from their formal specifications \cite{L08}. 
Coq uses two languages for proofs: {\em Gallina} (a pure functional programming language) for writing specification and {\em LTac} (a procedural language)
for the proof process manipulation.
Main success stories of the system are formalization of fully certified C-compiler \cite{Leroy, Blazy}, 
disjoint-set data structure \cite{Crary}, formalization of two way finite automata \cite{Doczkal}, multiplier circuit 
formalization \cite{Mohring} and coordination language \cite{Yi}. Main mathematical formalizations done in Coq include the formalization of 
Feit-Thompson theorem \cite{Gonthier1}, four-color theorem \cite{Gonthier2}, three gap theorem \cite{Mayero}, real analysis \cite{Boldo2} and 
theory of algebra \cite{Geuvers}. 
A new approach in Coq is presented in \cite{Tanaka18} that directly generates provably-safe C code.
Recently, Coq is used in the formal verification of dynamical systems \cite{Cohen17}, password quality checkers \cite{Ferreira17}, 
security protocol \cite{Palombo17}, QWIRE quantum circuit language \cite{Rand18}, complex data structure invariants \cite{Kaiser17}, 
component connectors \cite{ZHLS19, HNZYM} and the control function for the inverted pendulum \cite{Rouhling18}.
Moreover, a plug-in (called  SMTCoq) is developed in \cite{Ekici} to integrate external SMT solvers in Coq.
An IDE for integration of Coq projects into Eclipse is also developed in \cite{Faithfull18}. 

Compared to other ITPs, PVS is based on classical simple type theory and is without {\em proof-objects},
which allows all kinds of rewriting for numeric as well as symbolic equalities. 
It offers theory interpretation, dependent types, predicate sub-typing, powerful decision procedures, Latex 
support for specifications and proofs, and is user-friendly due to highly expressive specification language and powerful built-in automated deduction. It is also
integrated with other outside systems such as a BDD-based model checker and also serves as a back-end verification tool for computer algebra and code 
verification systems \cite{Maric}. 
During proof construction, PVS builds a graphical proof tree in which remaining proof obligations are at the leaves of tree.
Each node in the tree represents a sequence and each branch is considered as a (sub) proof 
goal that is followed from its off-spring branch with the help of a proof step.
PVS prover is based on \emph{sequent calculus} where each proof goal is a \emph{sequent} consisting of a sequence of formulas
called \emph{antecedents} and a sequence of formulas called {\em consequents}. The type system of PVS is not algorithmically decidable and theorem proving may be 
required to establish the type-consistency of a PVS specification. Theorems that need to be proved are called type-correctness conditions (TCCs). 
PVS is used in hardware and software verification \cite{Owre1, Kim}, concurrency problems verification \cite{Wim2}, 
file systems verification \cite{Wim3}, control systems \cite{Vitt}, cryptographic protocol \cite{Mauricio13},
microprocessor verification \cite{Srivas}, real time systems \cite{Shankar}, formalization of integral calculus 
\cite{Butler09} and medical devices \cite{Masci}.
Some recent work in PVS includes the specification of multi-window user interface \cite{Singh}, formalization of component connectors \cite{NMS19, Nawaz18}, 
analysis of distributed cognition systems \cite{Masci15}, genetic algorithms operators \cite{NM18bd} and cloud services \cite{Nawaz18b}.  
PVS along with its libraries is translated to the OMDoc/MMT framework in \cite{Kohlhase}. The proposed translation 
provides a universal representation format for theorem prover libraries.
Similarly, PVS is allowed in \cite{Gilbert} to export proof certificates that can be verified externally.
Moreover, denotational semantics for Answer Set Programming (ASP) is encoded in \cite{Aguado} and fundamental 
properties of ASP are proved with PVS theorem prover. 
Some of the differences between top ten famous proof-assistants are listed in Table \ref{itpcom}. 
\begin{table}[!ht]
 \centering
 \begin{threeparttable}

  \caption{Comparison of proof-assistants}\label{itpcom}\vspace*{-2mm}

  \begin{tabular}{|c|c|c|c|c|c|c|c|}
    \hline
    {\bf ITPs} & {\bf rel} & {\bf T} & {\bf dep. T} & {\bf dec. T}& {\bf state. $\mathbb{R}$} & {\bf rpif} & {\bf LLib} \\ \hline
     ACL2 & - & - & - & - & + & + & +\\ \hline
     Isabelle & ++ & + & - & + & + & + & + \\ \hline
     Metamath & + & - & - & - & + & + & + \\ \hline
     Twelf & + & + & + & + & - & - & - \\ \hline
     Agda & +++ & + & + & + & - & - & + \\ \hline
     Mizar & + & + & + & + & + & + & + \\ \hline
     HOL & ++ & + & - & + & + & - & + \\ \hline
     Nuprl & ++ & + & + & - & - & - & - \\ \hline
     Coq & + & + & + & + & + & - & +\\ \hline
     PVS & - & + & + & - & + & - & + \\ \hline    
     \end{tabular}
\begin{tablenotes}
 \item {\bf rel}: reliability, {\bf T}: typed, {\bf dep. T}: dependent type, {\bf dec. T}: decidable type, {\bf state. $\mathbb{R}$}: statement about $\mathbb{R}$, 
 {\bf rpif}: readable proof input files, {\bf LLib}: large library
\end{tablenotes}
\end{threeparttable}
\end{table}

\subsection{ATPs}
Isabelle is built around a relatively small core that implements many theories as classical FOL, constructive type theory, intuitionistic 
natural deduction and ZFC. Its meta-logic is based on the fragment of intuitionistic simple type theory 
that includes basic types as functional types. Whereas the terms are of typed $\lambda$-calculus. Only the type {\em prop} (proposition) is defined by the meta-logic 
and the formulas are terms of type {\em prop}. The meta-logic supports implication, the universal quantification and equality, and the inference rule is 
provided in natural-deduction style \cite{Maric}. For proofs, Isabelle combines HOL for writing specification and Isar as the language to describe procedures for proofs 
manipulation. Isabelle/HOL \cite{NPW02} is the most widely used system nowadays.  
Isabelle offers a rich infrastructure for high-level proof schemes. During theory development, both structured and unstructured proofs can be mixed freely. 
It is important to state that both HOL and Isabelle use {\em non-standard proof-objects} in the form of tactics for equational reasoning. 
This makes formalization relatively easy in both systems but it has the disadvantage that the {\em proof-objects} can not be used to see the proof details. 
In principal, both systems can be modified for {\em proof-objects} creation and storing.

HP used Isabelle in the design and analysis of the HP 9000 line of servers' Runway bus \cite{Camilleri}. 
The \emph{L4.verified project} at NICTA used Isabelle to prove the 
functional correctness of \emph{seL4} micro-kernel \cite{Klein2}. Moreover, Isabelle is successfully used in security protocols' 
correctness \cite{Butin}, formalization of Java programming language \cite{Oheimb}, Java virtual machine code soundness and 
completeness \cite{Klein1}, property verification of programming 
language semantics \cite{Klein3}.
A list of research projects that uses Isabelle can be found at: \url{isabelle.in.tum.de/community/projects}.
Recent works in Isabelle include the verification of Ethereum smart contract bytecode \cite{Amani18}, 
 imperative programs asymptotic time complexity verification \cite{Zhan18} and formalization of Akra-Bazzi method \cite{Eberl17},
deep learning theoretical foundations \cite{Bentkamp17}, Green's theorem \cite{Abdulaziz} and 
Markov chains and Markov decision processes with discrete time and 
discrete state-spaces \cite{Johannes17}.

From last 15 years, E theorem prover is constantly participating at CASC in more than one category (winnrer in SLH divison in CASC-27 (2019)). 
The semantics of E is purely decelrative and internal unique features are: shared terms with cached rewriting, folding feature vector 
indexing and fingerprint indexing. Unique features that are visible to the users are advanced and highly flexible search heuristics.
E main strengths are the generation of {\em proof-objects}, the automatic problem analysis and the support for the TPTP standrad for answers \cite{0001CV19}.
\emph{E} is successfully used in the reasoning of large ontologies \cite{Ramachandran}, software verification \cite{Ranise}
and certification \cite{Denney}. One of the main limitations in ATP is the lack of mechanism that allows proofs to guide the proof 
search for a new conjecture. In this regard, E is extended in \cite{Jakubuv16} with various new clause selection strategies. These strategies
are based on similarity of a clause with the conjecture. The use of watchlists (also known as hint list) in large E 
theories is explored in \cite{Goertzel18}, to improve the proof process automation.

Escher Verifier (the successor of Perfect Developer \cite{CR03}) is based on the Verified Design-by-Contract Paradigm \cite{Crocker14} inspired from 
Hoare logic and weakest precondition. It performs static analysis and formal verification of 
C programs by checking the out-of-bounds array indexing, arithmetic overflow, null-pointer de-referencing and other undefined behavior in the program. 
It extends the C language with additional keywords and constructs that are required in programs specifications expression and to give strength to the 
C type system \cite{Cartlon}. Term rewriting and FOL based theorem prover is implemented for the verification purpose. 
The unique feature of the tool is that it provides hints for the cases when the provers is unable to verify the code automatically \cite{HA11}. 
Escher verifier is used in the verification of C programs \cite{Crocker}, compilers \cite{Crocker3} and 
formal analysis of web applications \cite{Crocker2}.

Vampire \cite{Riazanov} is an ATP for FOL, based on equational superposition calculus and is one of the best ATPs at CASC. 
Unique features of Vampire include the generation of interpolants and implementation of a limited resource strategy (LRS). 
Moreover, symbol elimination is also implemented in Vampire that is used to automatically find first-order program properties.
It has a special mode for working with very large knowledge bases and can answer queries to them according to TPTP standars. On a multi-core system, Vampire can perform
several proof attempts in parallel \cite{KV13}.
The strength of ATPs such as Vampire, E and SPASS in proving theorems from MML is presented in \cite{Urban10}.
Some work on adding arithmetic to Vampire is done in \cite{Korovin11}.
Vampire is used in \cite{Grewe} to automate the soundness proofs of type systems.  
Vampire is also used for program analysis and in proving properties of loops with arrays \cite{Kovacs2}.
Cheap SAT solvers such as Avatar \cite{Reger} plays an important role in the success of Vampire.
In general, Vampire is well-suited in the domain of type soundness proofs. However, the use of Vampire 
relies heavily on the size of the chosen axiom set and on the concrete form of individual axioms.

Prover9 \cite{prover9-mace4}, the successor of the Otter prover, is a resolution based automated prover for equational logic and FOL. 
Main strength of Prover9 is that it is paired with Mace4. Users gives formulas and Prover9 attempts to find a proof. 
If proof is not find then Mace4 looks for a counter-example. Similalry, Prooftrans can be used to transform Prover9 proofs into more detailed proofs, 
simplify the justifications, re-number and expand proof steps, produce them in XML format, generate hints to guide subsequent searches and produces proofs for 
input to proof checkers such as IVY. 
Prover9 is used in the analysis of cryptographic-key Assignment schemes \cite{Eddin12}, verification of Alloy specification language 
\cite{Macedo12} and access control policies \cite{Sabri15}. Moreover some tasks in combinatorics on words \cite{Holub17} and
geometric procedure \cite{Padmanabhan13} is  also formalized in Prover9, along with proofs of theorems in Tarskian geometry \cite{Veroff15}.
Similarly, iProver \cite{Korovin08} is based on an instantiation framework for FOL called Inst-Gen \cite{Ganzinger03}.
First order reasoning is combined with ground reasoning in iProver with the help of SAT solver called MiniSat \cite{Niklas03}. 
Main strengths of iProver are: reasoning with large theories, a predicate elimination procedure as a preprocessing technique, 
EPR-based k-induction with counterexample, model representation with first order definitions in term algebra and 
proof extraction for resolution as well as instantiation. More details on iProver and other ATPs can be found in Appendix \ref{A1}.

Table \ref{diff6} compares the famous ATPs that perform best at CASC for some features. SonTPTP (Systems on TPTP) is an online interface for ATPs. It can be 
used by the users to run the ATP on TPTP (thousand problems for theorem provers) library or their own problems in the TPTP language \cite{S017}. 
It is important to point out that MaLARea \cite{Urban07} is not an ATP. It is a simple metasystem that combines several ATPs (E, SPASS, Vampire, etc) 
with a machine learning based component (SNoW system).  MaLARea interleaves the ATPs by first running them (in cycles) on problems, followed by machine learning 
from successful proofs. The learned information is then used to limit the set of axioms provided to ATPs in the next cycle. In CASC-J9 (2018), 
MaLARea comes first in the LTB division. 

\begin{table}[!ht]
 \centering
 \begin{threeparttable}
  \caption{Comparison of famous ATPs}\label{diff6}
  \begin{tabular}{|c|c|c|c|c|c|}
    \hline
    {\bf } & {\bf Type} & {\bf ILang} & {\bf Lib} & {\bf SS} & {\bf Web service}\\ \hline
     E & SB-FOP & C & - & + & SonTPTP\\ \hline
     Vampire & SB-FOP & C++ & + & - & SonTPTP\\ \hline    
     Prover9/Otter & RB-FOP & C & + & - & SonTPTP\\ \hline     
     SPASS & SB-FOP & Java/C & - & + & SonTPTP\\ \hline 
     Satallax & TB-HOP  & OCaml & - & + & SonTPTP \\ \hline 
     iProver & IB-FOP & OCaml & - & + & SonTPTP \\ \hline 
     LEO-II & RB-HOP & OCaml & + & + & SonTPTP \\ \hline 
     MaLARea & MS-ATP & Perl & + & - & - \\ \hline 
     \end{tabular}
\begin{tablenotes}
\item 
{\bf SS}: standalone system, {\bf FOP}: first-order prover 
 {\bf SB-FOP}: super-position-based FOP, {\bf RB-FOP}: resolution-based FOP, {\bf TB-FOP}: tableau-based FOP, {\bf TB-HOP}: tableau-based higher-order prover, 
 {\bf IB-FOP}: instantaition-based FOP, {\bf RB-HOP}: resolution-based HOP, {\bf SonTPTP}: systems on TPTP. {\bf MS-ATP}: metasystem for ATP
\end{tablenotes}
\end{threeparttable}
\end{table}

ATPs can be integrated (through hammers \cite{BPU16}) with ITPs for the proof automation in interactive proof development process. 
Hammers use ATPs to automatically find the proofs for user defined proof goals.
They combine the learning from previous proofs with translation of the goals to the logics of ATPs and reconstruction of the successfully found
proofs for the goals. Similarly, the SAT/SMT solvers can be used in ITPs by first translating and passing the goals to the fragment supported by a SAT/SMT solver. 
The SAT/SMT solver then solve the translated goal without human intervention and guidance. Table \ref{diff7} lists some of the hammers and SAT/SMT solvers that 
are developed and integrated with ITPs. Waldmeister \cite{Hillenbrand} is integrated with 
Agda in \cite{Foster}, but no SAT/SMT solvers is yet integrated with Agda. Moreover, PVS employs the Yices SMT solver as an oracle \cite{R06a}, 
but not integrated with any ATPs yet.
Some new theorem provers that aims to fill the gap between interactive and automated theorem proving such as Lean \cite{AMK19} 
 offers APIs to access features of SMT solvers (CVC4, Z3) and ATPs.
\begin{table}[!ht]
 \centering
  \caption{ATPs and SAT/SMT solvers for ITPs}\label{diff7}
 \begin{tabular}{|c|c|c|}
    \hline
    {\bf ITP} & {\bf Hammers} & {\bf SAT/SMT solvers} \\ \hline
     Isabelle/HOL & Sledgehammer \cite{Paulson2} & Yices in Isabelle/HOL \cite{EM08} \\ \hline
     HOL Light/HOL4 & HOLyHammer \cite{KKU15} & SMT solvers for HOL4 \cite{W011} \\ \hline
     Mizar & MizAR \cite{KU15a} & MiniSAT for Mizar \cite{N014} \\ \hline
     Coq & Hammer for Coq \cite{CK18} & SMTCoq \cite{Ekici}\\\hline
     ACL2 & ATPs for ACL2 \cite{JKU14} & Smtlink for ACL2 \cite{PengG15}\\\hline
     \end{tabular}
\end{table}\vspace*{-2mm}

\subsection{Some Future Directions}

Active research activity is going on in both ITPs and ATPs. 
Despite the great progress in last three decades, general purpose FOL based theorem provers are still unable to directly 
determine the satisfiability of a first-order formula. 
SMT problem deals with whether a formula written in first-order is satisfiable in some logical theory. 
One of the famous theorem prover for SMT problem is CVC4 \cite{Barrett11}. 
SMT solvers may not terminate on some problems due to undecidability of FOL.
In such cases, users would like to know the reason why the solver failed. 
Developing tools for SMT solvers which allows developers and users in helping the system to finish some proofs is an
interesting research area.

Currently, most of the SMT solvers display ``unknown'' when they are unable in proving the unsatisfiability of 
quantified formulas \cite{Reynolds117}. Main research direction in SMT solvers is to enable them to return counter models 
in case they fail to prove the unsatisfiability of quantified formulas that ranges from integers and inductive datatypes to free sorts. Popular SMT solvers (CVC4, Yices, Z3 etc) generally work in a sequential manner. One another research area is to parallelize SMT solving to better utilize the capability of hardware in multi-core systems. 
PZ3 \cite{CZSGS18} (the parallel solver for Z3) is one example for this kind of parallelization.
Another important area is the development of tools that can integrate SMT solvers in ITPs 
to increase the level of automation by offering safe tactics/strategies for solving proof goals automatically with the help of
external solvers. SMT solvers for ITPs listed in the preceding Table \ref{diff7} are some example of this. 

One of the main challenge in ATPs is reasoning with large ontologies, which are becoming more dominant
as large knowledge bases. Some techniques used for reasoning with large theories are based 
on methods for different axiom selection \cite{Hoder11, Sutcliffe07}. 
Machine learning is also used for axiom selection where previous knowledge is used to prove a 
conjecture \cite{Urban07, Urban08}. 
Framework for abstract refinement, where the axioms selection and reasoning phases are interleaved, can 
also be used in reasoning of large ontologies, as shown recently in  \cite{Hernandez17}.

Theorem provers have the limitations that it is not fast enough, the logic is inconvenient as a scripting
language and majority of theorem provers do not support graphics and visualization tools. Similarly, ITPs requires
heavy interactions between a user and the proof-assistant, which consumes a lot of time. 
IDE's in theorem provers , especially in ITPs, can substantially facilitate the creation of large proofs. 
However, very few of them are equipped with full-fledged IDE's.
Some future work in this direction includes: (i) making the provers fast and efficient, and (ii) development of IDE's and integration
of IDE's in different provers. This will make these tools more acceptable in industrial sector. 
Similarly, in ITPs, users make use of tactics that reduces a goal to simpler and smaller sub-goals. 
Another interesting area is the development of strategies/tactics by using
tactic languages, such as HITAC \cite{Aspinall} and Ltac \cite{Delahaye} which will allow users 
to elaborate proof strategies and combine tactics.

ITPs also lack the inter-operability among proof-assistants and other related tools, which means that tool support cannot be easily shared between ITPs. 
Translation of ITPs to a universal format is needed to overcome the duplication efforts
in the development of systems, their libraries and supporting periphery tools. Similarly,
some work \cite{Gastel, Gastel2} is done on integrating the model checking with theorem provers. 
However, integrating model checking with theorem proving is more difficult as it involves 
the mapping of models and mathematics involved in the analysis of the systems. 

The vision of QED manifesto is to develop a universal, computer-based database for all mathematical knowledge that is formalized
in logic and is supported by proofs that can be validated mechanically.
As shown in previous sections, theorem provers are diverse in nature with radically different foundations. 
On one hand, using various provers offers a diverse experience, which helps to better understand
the strengths and weaknesses of provers. However, on the other hand, the effort and overhead needed to learn even one prover 
effectively makes researchers to stick to using just one system. This results in duplication of similar work. 
A way of sharing the work and knowledge among provers would not be just appealing but it would also make 
provers more powerful and practical. One feasible approach is to import theorems from one prover to another.

Theorems between different systems are transported by translating the libraries between systems, as done in \cite{Obua06, Krauss10, Keller10}.
The main challenge in sharing theorems is to ensure a meaningful semantic match between the provers, meaning that logic, definitions, types and
treatment of functions, etc. in provers are compatible with each other.
Isabelle's sledgehammer \cite{Paulson2} describes the way for integrating different automation tools.
However, sledgehammer has number of limitations, such as unsound translation, primitive relevance filter and low performance on 
higher-order problems. Integrating ITPs with ATPs still requires a lot of research into approaches of interfacing. One
of the main challenge is a sound and reliable translation among different languages and logics.
Similarly, other main concern is the interpretation of ATP outputs back into ITP environment.

ITPs does have a large corpora of computer-understandable reasoning knowledge \cite{BHMN15, Harrison2} in the form of libraries. 
These corpora can play an important role in artificial intelligence based methods, such as concept matching, structure formation
and theory exploration. 
The ongoing fast progress in machine learning and data mining made it possible to use
these learning techniques on such corpora in guiding the proof search process, in proof automation and in developing proof tactics/strategies, as indicated in
recent works \cite{NMP19, Kaliszyk2, Gauthier17, Goertzel18, KUMO18}. 
Such machine learning systems can also be combined with ATPs on the large ITPs libraries to develop an artificial intelligence based feedback loops.
Another interesting area is to use evolutionary algorithms \cite{Back} in ITPs to find and evolve proofs. Till now,
effective search mechanisms for formal proofs are lacking and we believe that
evolutionary algorithms are more suitable for this task due to their suitability to solve search and optimization problems.
Moreover, investigating evolutionary/heuristic algorithms (as the program (proof) generator) and ITPs (as the proof verifier) to automatically find 
formal proofs for new conjectures is also worth pursuing. Some initial work can be found in \cite{Yang, Huang17}, where a GA is used with the Coq to automatically 
find formal proofs for theorems. 

A single tool based on machine learning is developed in \cite{Kaliszyk15} that can be used for every ITP on one condition: 
both the language and its corresponding library are available in a universal format, so that they can be easily put into the common selection 
algorithm. However, the universal format is generally infeasible and expensive for many applications. 
The reason is that it is very hard to built a universal format that can offer a good trade-off between universality
and simplicity. Even if such a universal format is available, the implementation of library export into the universal format is laborious. 

One of the such universal format for formal knowledge is the OMDoc/MMT framework \cite{Kohlhase06}, which has been used
to translate Mizar \cite{Iancu13}, HOL Light \cite{Kaliszyk14} and PVS \cite{Kohlhase} libraries into OMDoc/MMT framework . 
Their work has made the libraries becomes accessible to a wide range of OMDoc-based tools. 
It would be interesting to translate other famous ITPs logics and libraries to OMDoc/MMT framework for formal mathematics and knowledge management.
Also, translation of ITPs to one universal format will make the machine learning based premise selection to provers much easier.
Moreover, with flexible alignments \cite{Kaliszyk160} between the libraries, the developers of different provers can be guided in the 
approximate translation of contents across libraries and in reuse notations, such as to show one prover 
content in a form that looks familiar to other prover users.

\section{Conclusion} \label{s6}
Mechanical reasoning systems are actively developed since the birth of modern logic, and now these 
state-of-the-art tools are used in proving complicated mathematical theorems and verifying large computer systems.
A comprehensive survey on  mechanical reasoning systems (both ITPs and ATPs) is presented in this work. Main characteristics, strengths, differences 
and application areas of the systems are investigated. Some future research directions based on recent work are also discussed.
In summary, we find that formalization with theorem provers have not become mature enough to adopt the working style of vast  mathematical 
community. It still needs: (i) better libraries support for proofs automation, (ii) better ways of knowledge sharing between the proof systems, 
(iii) computation incorporation and verification, (iv) better means to store and search the background facts, (v) improved interfaces and integrations of  
ATPs, ITPs and SAT/SMT solvers, and (vi) better support for the machine learning and deep mining techniques for proof guidance and automation. 
Currently, a wider researcher community is working on these problems and it is sanguinely estimated that mechanically formalized and verified 
mathematics will be a commonplace till the mid of this century. 

In this survey, judgments which we made about theorem provers may be subjective. The authors hope that this survey will provide a quick and easy 
guide to the interested users and people without good mathematics knowledge into the work of theorem provers. 
In advance, we allege for misrepresentation of these systems from 
their perspective developers or owners. We feel happy to be informed  via e-mail about any lapses at \texttt{msaqibnawaz@hit.edu.cn}.

\section*{Acknowledgments}
The work has been supported by the National Natural Science Foundation of China under grant no. 61772038, 61532019 and 61272160, and
the Guandong Science and Technology Department (Grant no. 2018B010107004).
\bibliographystyle{abbrv}
\small{
\bibliography{sb}
}


%

\appendix \label{A1}
System details of 27 theorem provers.
\begin{table}[ht]
 \centering 
 \rotatebox{90}{
\scalebox{0.95}{
\begin{tabular}{|p{0.02\textwidth}|p{0.1\textwidth}|p{4cm}|p{4cm}|p{3.9cm}|p{4.57cm}|}
\hline
\parbox[t]{2mm}{\multirow{4}{*}{\rotatebox[origin=c]{90}{\textbf{General}}}} & \textbf{Name} & \textbf{HR} & \textbf{ICS} & \textbf{Analytica} & \textbf{Zenon} \\
 & \textbf{Contributor} & Simon Colton & Jean-Christophe Fillitre & E. Clarke \& X. Zhao & R. Bonichon, D. Delahaye \& D. Doligez \\
 & \textbf{1st Rel} & 2002 & 2002 & 1990 & 2007\\
 & \textbf{Ind/Uni/Ind} & University of Edinburgh & SRI International, USA & Carnegie Mellon University & Independent\\ \hline
\parbox[t]{2mm}{\multirow{11}{*}{\rotatebox[origin=c]{90}{\textbf{Implementation}}}} & \textbf{CLang} & Java & Ocaml & Mathematica & OCaml\\
 & \textbf{Prog.P} & Functional, Concurrent & Functional, Imperative, OO & Procedural, Functional, OO & Functional, Imperative, OO\\
 &\textbf{LV}  & HR 2.0 & Yices & Analytica 2 & 0.8.2\\
 &\textbf{LT} & Open source & Open source & Free BSD & BSD and MIT \\
 &\textbf{UI}  & CLI & CLI & GUI & CLI\\
 & \textbf{OS} & Linux, Windows & Linux, Solaris, MAC & Cross & Cross  \\
 & \textbf{Lib} & API & No & Standard library & Standard library \\
 & \textbf{CG} & No & No & No & No \\
 & \textbf{Ed} & Yes & Yes  & Yes & No \\
 & \textbf{Ext} & Yes & Yes &  Yes & Yes\\
 &\textbf{I/O}  & No & No & Tex files & No \\ \hline
\parbox[t]{2mm}{\multirow{6}{*}{\rotatebox[origin=c]{90}{\textbf{Logico-Math}}}} & \textbf{TType} & Theorem generator & Decision procedure & ATP & 
ATP ({\tiny Algebraic specification \& proof system})\\
 & \textbf{CLogic} & FOL & FOL with equality \& quantifier free & FOL & FOL (with polymorphic \& equality) \\
 & \textbf{TV} & Binary & Binary & Binary & Binary\\
 & \textbf{ST} & No & No & No & B-Method set theory\\
 & \textbf{Calculus} & Inductive & Deductive & Deductive & Deduction modulo (Tableau method)\\
 & \textbf{ProofKernel} & No & No & Yes & No \\ \hline
\parbox[t]{2mm}{\multirow{3}{*}{\rotatebox[origin=c]{90}{\textbf{Others}}}} & \textbf{App. Areas} & Produce large number of theorems for 
        testing ATP Systems & Embedded in application to provide deductive services & 
	      Symbolic computation system & Used in focal environment, Object oriented algebra specification \\
 &\textbf{Eval}  & Zariski Specification & NASA, part of PVS & Policy analysis & TPTP Category Set (227 out of 462)
 SEU (110 out of 900) \\
 & \textbf{Unique Features} & Machine Learning  & API for proof search and symbolic simulation & Translated to OMDoc framework & Produce low 
            level proof directly for Coq\\
\hline
\end{tabular}
}
}
\label{ACL2}
\end{table}

\begin{table}[ht]
 \centering 
  \rotatebox{90}{
\begin{tabular}{|p{0.02\textwidth}|p{0.1\textwidth}|p{4.2cm}|p{4.2cm}|p{4.2cm}|p{4.2cm}|}
\hline
\parbox[t]{2mm}{\multirow{4}{*}{\rotatebox[origin=c]{90}{\textbf{General}}}} & \textbf{Name} & \textbf{Yarrow} & \textbf{Watson} & \textbf{KRHyper/E-KRHyper/Hyper} & \textbf{iProver} \\
 & \textbf{Contributor} & Jan Zwanenburg & M. Randall Holmes & Bj\"{o}rn Pelzer & Konstantin Korovin\\
 & \textbf{1st Rel} & 1997 & 2006 & 2007 & 2008\\
 & \textbf{Ind/Uni/Ind} & Eindhoven University & Boise State University & Koblenz University & University of Manchester\\ \hline
\parbox[t]{2mm}{\multirow{11}{*}{\rotatebox[origin=c]{90}{\textbf{Implementation}}}} & \textbf{CLang} & Haskell & Standard ML & OCaml & OCaml\\
 & \textbf{Prog.P} & Functional, Modular & Functional and Imperative & Functional, Imperative, OO & Functional, Imperative, OO \\
 & \textbf{LV} & V1.20 & 0.8.2 & 1.4 & V0.99\\
 & \textbf{LT} & Free BSD & BSD and MIT GNU General & GNU GPL & GNU\\
 & \textbf{UI} & GUI \& CLI & CLI & CLI & CLI\\
 & \textbf{OS} & Unix, Linux & Linux & Windows and Unix & Linux \\
 & \textbf{Lib} & Fudget for graphical interface & No & No & No \\
 & \textbf{CG} & No & No & No & No\\
 & \textbf{Ed} & Yes & No & Yes & Yes\\
 & \textbf{Ext} & Yes & Yes & Yes & Yes\\
 & \textbf{I/O} & Polymorphic & No & TPTP supported protein format & No \\ \hline
\parbox[t]{2mm}{\multirow{5}{*}{\rotatebox[origin=c]{90}{\textbf{Logico-Math}}}} &\textbf{TType} & ITP & ITP & ATP and model generator & ATP\\
 & \textbf{CLogic} & Constructive & HOL & FOL with equality & FOL\\
 & \textbf{TV}&  Binary & Binary & Binary & Binary \\
 & \textbf{ST} & No & Quine's set theory & No & No\\
 &\textbf{Calculus} & Typed $\lambda$-calculus & Typed $\lambda$-calculus  & Hyper tableau calculus & Instantiation calculus\\
 & \textbf{ProofKernel} & Yes & No & No & No\\ \hline
\parbox[t]{2mm}{\multirow{3}{*}{\rotatebox[origin=c]{90}{\textbf{Others}}}} & \textbf{App. Areas} & Representation environment for logics 
\& programming languages. & Software verification, 
       model checking, education \& mathematics & Embedded in knowledge representation systems & Hardware verification and finite modeling\\
 & \textbf{Eval} & Formalized type theory & TPTP category  & Solved 74\% of the subest of TPTP & CASC-26 EPR division winner\\
& \textbf{Unique Features} & Experiment context for testing pure type system & Type free HOL support
 & Used for description logic problems & Modular combination of proposition and instantiational reasoning\\
\hline
\end{tabular}
}
\label{Yarrow}
\end{table}

\begin{table}[ht]
 \centering 
  \rotatebox{90}{
\begin{tabular}{|p{0.02\textwidth}|p{0.1\textwidth}|p{4.2cm}|p{4.2cm}|p{4.2cm}|p{4.2cm}|}
\hline
\parbox[t]{2mm}{\multirow{4}{*}{\rotatebox[origin=c]{90}{\textbf{General}}}} & \textbf{Name} & \textbf{JAPE} & \textbf{E-Darwin} & \textbf{leanCoP} & \textbf{LEO-II}\\
 & \textbf{Contributor} & Richard Bornat & Baumgartner Otten & Jens Otten & C. Benzmuller, F. Theiss, N. Sultana\\
 & \textbf{1st Rel} & 1996 & 2005 & 2003 & 2012\\
 & \textbf{Ind/Uni/Ind} & Queen Marry, University of London & Koblenz University & University of Oslo & 
                     Freie University Berlin and Cambridge University\\ \hline
\parbox[t]{2mm}{\multirow{11}{*}{\rotatebox[origin=c]{90}{\textbf{Implementation}}}} & \textbf{CLang} & Java & OCaml & Prolog & OCaml\\
 & \textbf{Prog.P} & Object oriented (OO)  & Functional, imperative, OO & Logic programming & Functional, Imperative, OO \\
 & \textbf{LV} & v7-d15 & 1.5 & 2.1  & 1.7 2015\\
 & \textbf{LT} & GNU GPL & GNU General  & GNU general & BSD\\
 & \textbf{UI} & GUI & CLI & CLI & CLI\\
 & \textbf{OS} & Linux, Mac & Unix, Windows & Windows, Unix, Linux, Mac & Unix, Windows\\
 & \textbf{Lib} & No  & No & No & Yes\\
 & \textbf{CG} & No & No & No & No\\
 & \textbf{Ed} & Yes & Yes & Yes & Yes \\
 & \textbf{Ext} & Yes & Yes & Yes & Yes \\
 & \textbf{I/O} & No & Input TPTP or TME format  & LeanCoP or TPTP syntax & TPTP THF language\\ \hline
\parbox[t]{2mm}{\multirow{5}{*}{\rotatebox[origin=c]{90}{\textbf{Logico-Math}}}} & {\bf TType} & Proof assistant & ATP & ATP & ATP+ITP\\
 & \textbf{CLogic} & FOL & FOL clausal with equality & First-order intuitionistic & HOL\\
 & \textbf{TV} & Binary & Binary & Binary & Binary\\
 & \textbf{ST} & Yes & No & No & No\\
 & \textbf{Calculus} & Deductive \& Sequent calculus & Model evaluation  & Connection/tableau calculus & Resolution by unification and equality (RUE)\\
 & \textbf{ProofKernel} & No & No & No & No\\ \hline
\parbox[t]{2mm}{\multirow{3}{*}{\rotatebox[origin=c]{90}{\textbf{Others}}}} & \textbf{App. Areas} & Used as a proof assistant and implement JAPE theories & Encrypt and solve problems & Formalization & Cooperation with first-order ATP\\
 & \textbf{Eval} & Teaching purpose tool & EPR winner at CASC-20 and J3 &  Third in FOF division at CASC-22 & CASC-J5 winner in THF division \\
&\textbf{Unique Features} & Forward reasoning and logic encoding & Back-jumping and dynamic backtracking & Program can be easily modified for specific task or application due to its compact code & Cooperative Proof Search
 \\
\hline
\end{tabular}
}
\label{CVC4, E-Darwin, leanCoP and LEO-II}
\end{table}

\begin{table}[ht]
 \centering 
  \rotatebox{90}{
\begin{tabular}{|p{0.02\textwidth}|p{0.1\textwidth}|p{4.2cm}|p{4.2cm}|p{4.2cm}|p{4.2cm}|}
\hline
\parbox[t]{2mm}{\multirow{4}{*}{\rotatebox[origin=c]{90}{\textbf{General}}}} & \textbf{Name} & \textbf{MaLARea} & \textbf{Muscadet} & \textbf{Princess} & \textbf{Satallax}\\
 & \textbf{Contributor} & Josef Urban & Dominique Pastre & Philipp Rummer & Chad E. Brown\\
 & \textbf{1st Rel} & 2007 & 2003 & 2008 & 2010\\
 & \textbf{Ind/Uni/Ind} & Charles University & University Paris Descartes & Uppsala university & Saarland University\\ \hline
\parbox[t]{2mm}{\multirow{11}{*}{\rotatebox[origin=c]{90}{\textbf{Implementation}}}} & \textbf{CLang} & Perl & SWI-Prolog & Scala & OCaml \\
 & \textbf{Prog.P} & Functional, Imperative, OO & Logic programming & OO, functional, concurrent & Functional, imperative, OO\\
 & \textbf{LV} & 0.5  & 4.5  & V2.1 & 3.0\\
 & \textbf{LT} & GPL2 & BSD & LGPL  & BSD\\
 & \textbf{UI} & CLI & CLI & CLI & CLI\\
 & \textbf{OS} & Linux & Unix, Linux  & Linux & Linux\\
 & \textbf{Lib} & No & No & No & No\\
 & \textbf{CG} & No & No & No & No\\
 & \textbf{Ed} & Yes & Yes & Yes & Yes \\
 & \textbf{Ext} & Yes & Yes & Yes & Yes\\
 & \textbf{I/O} & No & No & Native language SMT & No \\ \hline
\parbox[t]{2mm}{\multirow{5}{*}{\rotatebox[origin=c]{90}{\textbf{Logico-Math}}}} & \textbf{TType} & ATP & Knowledge base TP & Theorem prover & ATP\\
 & \textbf{CLogic} & HOL & second-order logic & FOL & HOL \\
 & \textbf{TV} & Binary & Binary & Binary & Binary\\
 & \textbf{ST} & No & No & Modular linear integer arithmetic & Church's simple type theory\\
 & \textbf{Calculus} & Deductive & Natural deduction  & Free variable tableau,Constrained sequent calculus & Tableau calculus\\
 & \textbf{ProofKernel} & No & No & No & No\\ \hline
\parbox[t]{2mm}{\multirow{3}{*}{\rotatebox[origin=c]{90}{\textbf{Others}}}} & \textbf{App. Areas} & Learning and reasoning system for proving in large formal libraries & Topological linear spaces, cellular automata & Software verification and model checking & Formalization
\\
 & \textbf{Eval} & CASC-24 LTB division winner &  Winner CASC-JC in IJCAR 2001 & Solved problems from QF-LIA category of SMT Library & CASC-26 THF division winner \\
& \textbf{Unique Features} & Utilize ATP as core system with AI techniques. &
 Efficient for the problems containing too many axioms and formulas & Solved quantified modulo linear integer arithmetic & 
 Semantic embedding and cut simulation
 \\
\hline
\end{tabular}
}
\label{MaLARea and Muscadet}
\end{table}

\begin{table}[ht]
 \centering 
  \rotatebox{90}{
\begin{tabular}{|p{0.02\textwidth}|p{0.1\textwidth}|p{4.2cm}|p{4.2cm}|p{4.2cm}|p{4.2cm}|}
\hline
\parbox[t]{2mm}{\multirow{4}{*}{\rotatebox[origin=c]{90}{\textbf{General}}}} & \textbf{Name} & \textbf{SPASS} & \textbf{Coral} & \textbf{DISCOUNT} & \textbf{DORIS} \\
 & \textbf{Contributor} & Christophe Weidenbach & Alan Bundy, Graham Steel and Monika Maidl & J\"{o}rg Denzinger & Johan Bos\\
 & \textbf{1st Rel} & 1999 & 2006 & 1997 & 1998\\
 & \textbf{Ind/Uni/Inde} &  Max Planck Institute for Computer Science   & University of Edinburgh & University of Calgary & University of Edinburgh\\ \hline
\parbox[t]{2mm}{\multirow{11}{*}{\rotatebox[origin=c]{90}{\textbf{Implementation}}}} & \textbf{CLang} & Java & Built on SPASS theorem prover & C & Prolog\\
 & \textbf{Prog.P} & Functional & Functional  & Procedural & Logic programming\\
 & \textbf{LV} & 3.9 & 2008 & 2.0  & DORIS 2001\\
 & \textbf{LT} & Free BSD  & Free BSD  & Free BSD  & Free BSD \\
 & \textbf{UI} & CLI+GUI & GUI & CLI & CLI \\
 & \textbf{OS} & Windows, MAC, Linux &  Cross & Linux, Solaris & Cross Platform  \\
 & \textbf{Lib} & Yes & Yes & Yes & No \\
 & \textbf{CG} & No & No & No & No\\
 & \textbf{Ed} & Yes & Yes & Yes & Yes \\
 & \textbf{Ext} & Yes & Yes & No & Yes \\
 & \textbf{I/O} & No & No & E(Universal Implication) & No  \\ \hline
\parbox[t]{2mm}{\multirow{5}{*}{\rotatebox[origin=c]{90}{\textbf{Logico-Math}}}} & \textbf{TType} & ATP & Inductive theorem prover & 
Distributed equational TP & TP+ Semantic analyzer\\
 & \textbf{CLogic} & FOL & FOL & FOL & FOL\\
 & \textbf{TV} & Binary & Binary & Binary & Binary \\
 & \textbf{ST} & No & No & No & No \\
 & \textbf{Calculus} & Superposition calculus & Tableau calculus  & Pure unit equality & Lambda calculus\\
 & \textbf{ProofKernel} & No & No & Yes & No\\ \hline
\parbox[t]{2mm}{\multirow{3}{*}{\rotatebox[origin=c]{90}{\textbf{Others}}}} & \textbf{App. Areas} & Analysis of security protocols, collision avoidance protocols & Cryptographic security protocol analysis, find attacks on faulty security protocols  & Machine learning & Computational semantics, cover various linguistic phenomenas\\
 & \textbf{Eval} & Discover new attacks on the Asokan-Ginzboorg protocol & Discover new attacks on the Taghidri and Jackson improved protocol & Entrance competition Discount/GL & Study behavior of ROB's algorithm
   \\
& \textbf{Unique Features} & ATP with equality & Find counterexample to inductive conjecture & Based on teamwork method for knowledge based distribution & Translate English text into discourse representation structure
 \\
\hline
\end{tabular}
}
\label{SPASS and Coral}
\end{table}

\begin{table}[ht]
 \centering 
  \rotatebox{90}{
\begin{tabular}{|p{0.02\textwidth}|p{0.1\textwidth}|p{4.2cm}|p{4.2cm}|p{4.2cm}|p{4.2cm}|}
\hline
\parbox[t]{2mm}{\multirow{4}{*}{\rotatebox[origin=c]{90}{\textbf{General}}}} & \textbf{Name} & \textbf{Getfol} & \textbf{Goedel} & \textbf{Expander} & \textbf{Geometry Expert}\\
 & \textbf{Contributor} & Fausto Giunchiglia & Johan Belinfante & Peter Padawitz & Xiao-Shan Gao\\
 & \textbf{1st Rel} & 1994 & 2005 & 2007 & 1998\\
 & \textbf{Ind/Uni/Ind} & University of Trento & Georgia Institute of Technology & Dortmund university & Key Laboratory of China\\ \hline
\parbox[t]{2mm}{\multirow{11}{*}{\rotatebox[origin=c]{90}{\textbf{Implementation}}}} & \textbf{CLang} & Common Lisp & Mathematica & O'Haskell ( extention of Haskell) & Java \\
 & \textbf{Prog.P} & Meta reflective, OO, functional & Functional, procedural & Concurrent programming & Functional \\
 & \textbf{LV} & 2.001 & 2014 & Expander & MMP/Geometer\\
 & \textbf{LT} &  Free BSD & Free BSD & Free BSD  & GNU general public liscence\\
 & \textbf{UI} & GUI & GUI & GUI & GUI \\
 & \textbf{OS} & Unix & Cross & Cross & Cross   \\
 & \textbf{Lib} & Yes & No & Yes & No \\
 & \textbf{CG} & Yes & No & No & No \\
 & \textbf{Ed} & Yes & Yes & Yes & Yes\\
 & \textbf{Ext} & Yes & Yes & Yes & Yes \\
 & \textbf{I/O} & Proof-script & No & No & No \\ \hline
\parbox[t]{2mm}{\multirow{5}{*}{\rotatebox[origin=c]{90}{\textbf{Logico-Math}}}} & \textbf{TType} & ITP & ATP & ATP & Automatic Geometric TP \\
 & \textbf{CLogic} & FOL & FOL & FOL & Dynamic logic models\\
 & \textbf{TV} & Binary & Binary & Binary & Binary \\
 & \textbf{ST} & No & ZF set theory & Swinging types & No\\
 & \textbf{Calculus} & Natural deduction & Natural deduction  & Narrow \& Fixed point co-induction & Euclidean and differential geometry\\
 & \textbf{ProofKernel} & No & No & No & No \\ \hline
\parbox[t]{2mm}{\multirow{3}{*}{\rotatebox[origin=c]{90}{\textbf{Others}}}} & \textbf{App. Areas} & In various data structure real world embedding 
& Derive new theorem for automated reasoning & Testing algebraic data type and functional logic program & Teaching geometry, algebra and physics in China at school \\
   & \textbf{Eval} & Entrance in CADE-17 & 393 example of QAIF in 91 seconds & Super concentrators & Wu's method implemented
   \\
& \textbf{Unique Features} & Meta-theory implementation & Reduce number of steps & Interactive term rewriting, graph transformation, 
several representation of formal expression & Automated geometric diagram construction
 \\
\hline
\end{tabular}
}
\label{Getfol and Goedel}
\end{table}

\begin{table}[ht]
 \centering 
  \rotatebox{90}{
\begin{tabular}{|p{0.02\textwidth}|p{0.1\textwidth}|p{4.2cm}|p{4.2cm}|p{4.2cm}|}
\hline
\parbox[t]{2mm}{\multirow{4}{*}{\rotatebox[origin=c]{90}{\textbf{General}}}} & \textbf{Name} & \textbf{DTP} & \textbf{Z/EVES} & \textbf{Graffiti} \\
 & \textbf{Contributor} & Don Geddis & Mark Saaltinkz & Epstein\\
 & \textbf{1st Rel} & 1995 & 1997 & 1986 \\
 & \textbf{Ind/Uni/Ind} & Stanford University & University of Kent & University of Houston \\ \hline
\parbox[t]{2mm}{\multirow{11}{*}{\rotatebox[origin=c]{90}{\textbf{Implementation}}}} & \textbf{CLang} & Common Lisp & Common Lisp & C++ \\
 & \textbf{Prog.P} & Meta reflective, OO, functional & Meta reflective, OO, functional, procedural & Procedural, OO, statically type, type checking \\
 & \textbf{LV} & 3.0 & 2.4.1 & GRAFFITI \\
 & \textbf{LT} & Free BSD & Free BSD & Free BSD  \\
 & \textbf{UI} & CLI & GUI/CLI & GUI \\
 & \textbf{OS} & Cross & Windows, Unix, Linux, Mac & Uinx \\
 & \textbf{Lib} & Yes Epilog & Yes & Yes \\
 & \textbf{CG} & No & Yes & No \\
 & \textbf{Ed} & Yes & Yes & Yes \\
 & \textbf{Ext} & Yes & Yes & Yes \\
 & \textbf{I/O} & KIF (Knowledge Interchange Format) & Latex & No \\ \hline
\parbox[t]{2mm}{\multirow{5}{*}{\rotatebox[origin=c]{90}{\textbf{Logico-Math}}}} & \textbf{TType} & Modal elimination theorem prover & ATP+ITP & Decision procedure \\
 & \textbf{CLogic} & FOL & FOL & Graph theory \\
 & \textbf{TV} & Binary  & Binary & Binary \\
 & \textbf{ST} & Horn theories & ZF set theory, Axiomatic set theory & Yes \\
 & \textbf{Calculus} & First-order predicate calculus & $\lambda$-calculus & Deductive \\
 & \textbf{ProofKernel} & No & No & No \\ \hline
\parbox[t]{2mm}{\multirow{3}{*}{\rotatebox[origin=c]{90}{\textbf{Others}}}} & \textbf{App. Areas} & Black box inference engine for various machine learning program & General theorem prover  & Used in chemistry and mathematics (graph theory) for making conjecture \\
 & \textbf{Eval} & & Specification & Barbara Project \\
& \textbf{Unique Features} & Domain independent control of inference & Theorem prover, syntax and type checker, domain checker & Pedagogic tool \\
\hline
\end{tabular}
}
\label{DTP and Z/EVES}
\end{table}

\end{document}